\pdfoutput=1
\documentclass[12pt]{article}

\usepackage{graphicx}
\usepackage{enumitem}   
\usepackage{float}
\input{epsf}

\advance \topmargin by -\headheight
\advance \topmargin by -\headsep

\evensidemargin \oddsidemargin
\begin{document}

\topmargin -.6in

\def\rh{{\hat \rho}}
\def\alie{{\hat{\cal G}}}
\newcommand{\sect}[1]{\setcounter{equation}{0}\section{#1}}
\renewcommand{\theequation}{\thesection.\arabic{equation}}

\def\rf#1{(\ref{eq:#1})}
\def\lab#1{\label{eq:#1}}
\def\nonu{\nonumber}
\def\br{\begin{eqnarray}}
\def\er{\end{eqnarray}}
\def\be{\begin{equation}}
\def\ee{\end{equation}}
\def\eq{\!\!\!\! &=& \!\!\!\! }
\def\foot#1{\footnotemark\footnotetext{#1}}
\def\lb{\lbrack}
\def\rb{\rbrack}
\def\llangle{\left\langle}
\def\rrangle{\right\rangle}
\def\blangle{\Bigl\langle}
\def\brangle{\Bigr\rangle}
\def\llbrack{\left\lbrack}
\def\rrbrack{\right\rbrack}
\def\lcurl{\left\{}
\def\rcurl{\right\}}
\def\({\left(}
\def\){\right)}
\newcommand{\nit}{\noindent}
\newcommand{\ct}[1]{\cite{#1}}
\newcommand{\bi}[1]{\bibitem{#1}}
\def\lskip{\vskip\baselineskip\vskip-\parskip\noindent}
\relax

\def\tr{\mathop{\rm tr}}
\def\Tr{\mathop{\rm Tr}}
\def\trace{\widehat{\rm Tr}}
\def\v{\vert}
\def\bv{\bigm\vert}
\def\Bgv{\;\Bigg\vert}
\def\bgv{\bigg\vert}
\newcommand\partder[2]{{{\partial {#1}}\over{\partial {#2}}}}
\newcommand\funcder[2]{{{\delta {#1}}\over{\delta {#2}}}}
\newcommand\Bil[2]{\Bigl\langle {#1} \Bigg\vert {#2} \Bigr\rangle}  
\newcommand\bil[2]{\left\langle {#1} \bigg\vert {#2} \right\rangle} 
\newcommand\me[2]{\left\langle {#1}\bv {#2} \right\rangle} 
\newcommand\sbr[2]{\left\lbrack\,{#1}\, ,\,{#2}\,\right\rbrack}
\newcommand\pbr[2]{\{\,{#1}\, ,\,{#2}\,\}}
\newcommand\pbbr[2]{\lcurl\,{#1}\, ,\,{#2}\,\rcurl}

\def\ket#1{\mid {#1} \rangle}
\def\bra#1{\langle {#1} \mid}
\newcommand{\braket}[2]{\langle {#1} \mid {#2}\rangle}
%
\def\a{\alpha}
\def\at{{\tilde A}^R}
\def\atc{{\tilde {\cal A}}^R}
\def\atcm#1{{\tilde {\cal A}}^{(R,#1)}}
\def\b{\beta}
\def\dc{{\cal D}}
\def\d{\delta}
\def\D{\Delta}
\def\eps{\epsilon}
\def\vareps{\varepsilon}
\def\g{\gamma}
\def\G{\Gamma}
\def\grad{\nabla}
\def\h{{1\over 2}}
\def\l{\lambda}
\def\L{\Lambda}
\def\m{\mu}
\def\n{\nu}
\def\o{\over}
\def\om{\omega}
\def\O{\Omega}
\def\p{\phi}
\def\P{\Phi}
\def\pa{\partial}
\def\pr{\prime}
\def\pt{{\tilde \Phi}}
\def\qs{Q_{\bf s}}
\def\ra{\rightarrow}
\def\s{\sigma}
\def\S{\Sigma}
\def\t{\tau}
\def\th{\theta}
\def\Th{\Theta}
\def\tpp{\Theta_{+}}
\def\tmm{\Theta_{-}}
\def\tpg{\Theta_{+}^{>}}
\def\tms{\Theta_{-}^{<}}
\def\tp0{\Theta_{+}^{(0)}}
\def\tm0{\Theta_{-}^{(0)}}
\def\ti{\tilde}
\def\wti{\widetilde}
\def\jc{J^C}
\def\bj{{\bar J}}
\def\sj{{\jmath}}
\def\bsj{{\bar \jmath}}
\def\bp{{\bar \p}}
\def\vp{\varphi}
\def\ve{\varepsilon}
\def\vt{{\tilde \varphi}}
\def\faa{Fa\'a di Bruno~}
\def\ca{{\cal A}}
\def\cb{{\cal B}}
\def\ce{{\cal E}}
\def\cg{{\cal G}}
\def\cgh{{\hat {\cal G}}}
\def\ch{{\cal H}}
\def\chh{{\hat {\cal H}}}
\def\cl{{\cal L}}
\def\cm{{\cal M}}
\def\cn{{\cal N}}
\def\u2{\mid u\mid^2}
\def\ub{{\bar u}}
\def\z2{\mid z\mid^2}
\def\zb{{\bar z}}
\def\w2{\mid w\mid^2}
\def\wb{{\bar w}}
\newcommand\sumi[1]{\sum_{#1}^{\infty}}   
\newcommand\fourmat[4]{\left(\begin{array}{cc}  
{#1} & {#2} \\ {#3} & {#4} \end{array} \right)}

%
\def\lie{{\cal G}}
\def\kmlie{{\hat{\cal G}}}
\def\dlie{{\cal G}^{\ast}}
\def\elie{{\widetilde \lie}}
\def\edlie{{\elie}^{\ast}}
\def\hlie{{\cal H}}
\def\flie{{\cal F}}
\def\wlie{{\widetilde \lie}}
\def\f#1#2#3 {f^{#1#2}_{#3}}
\def\winf{{\sf w_\infty}}
\def\win1{{\sf w_{1+\infty}}}
\def\hwinf{{\sf {\hat w}_{\infty}}}
\def\Winf{{\sf W_\infty}}
\def\Win1{{\sf W_{1+\infty}}}
\def\hWinf{{\sf {\hat W}_{\infty}}}
\def\Rm#1#2{r(\vec{#1},\vec{#2})}          
\def\OR#1{{\cal O}(R_{#1})}           
\def\ORti{{\cal O}({\widetilde R})}           
\def\AdR#1{Ad_{R_{#1}}}              
\def\dAdR#1{Ad_{R_{#1}^{\ast}}}      
\def\adR#1{ad_{R_{#1}^{\ast}}}       
\def\KP{${\rm \, KP\,}$}                 
\def\KPl{${\rm \,KP}_{\ell}\,$}         
\def\KPo{${\rm \,KP}_{\ell = 0}\,$}         
\def\mKPa{${\rm \,KP}_{\ell = 1}\,$}    
\def\mKPb{${\rm \,KP}_{\ell = 2}\,$}    
%
\def\rlx{\relax\leavevmode}
\def\inbar{\vrule height1.5ex width.4pt depth0pt}
\def\IZ{\rlx\hbox{\sf Z\kern-.4em Z}}
\def\IR{\rlx\hbox{\rm I\kern-.18em R}}
\def\IC{\rlx\hbox{\,$\inbar\kern-.3em{\rm C}$}}
\def\IN{\rlx\hbox{\rm I\kern-.18em N}}
\def\IO{\rlx\hbox{\,$\inbar\kern-.3em{\rm O}$}}
\def\IP{\rlx\hbox{\rm I\kern-.18em P}}
\def\IQ{\rlx\hbox{\,$\inbar\kern-.3em{\rm Q}$}}
\def\IF{\rlx\hbox{\rm I\kern-.18em F}}
\def\IG{\rlx\hbox{\,$\inbar\kern-.3em{\rm G}$}}
\def\IH{\rlx\hbox{\rm I\kern-.18em H}}
\def\II{\rlx\hbox{\rm I\kern-.18em I}}
\def\IK{\rlx\hbox{\rm I\kern-.18em K}}
\def\IL{\rlx\hbox{\rm I\kern-.18em L}}
\def\one{\hbox{{1}\kern-.25em\hbox{l}}}
\def\0#1{\relax\ifmmode\mathaccent"7017{#1}%
B        \else\accent23#1\relax\fi}
\def\omz{\0 \omega}
%
\def\ltimes{\mathrel{\vrule height1ex}\joinrel\mathrel\times}
\def\rtimes{\mathrel\times\joinrel\mathrel{\vrule height1ex}}
%
\def\mark{\noindent{\bf Remark.}\quad}
\def\prop{\noindent{\bf Proposition.}\quad}
\def\theor{\noindent{\bf Theorem.}\quad}
\def\name{\noindent{\bf Definition.}\quad}
\def\exam{\noindent{\bf Example.}\quad}
\def\proof{\noindent{\bf Proof.}\quad}

\begin{titlepage}
\vspace*{-1cm}

\vskip 3cm

\vspace{.2in}
\begin{center}
{\large\bf  A Quasi Self-Dual Skyrme Model}
\end{center}

\vspace{.5cm}

\begin{center}
L. A. Ferreira$^{\dagger,}$\footnote{laf@ifsc.usp.br} and L. R. Livramento$^{\dagger ,\star ,}$\footnote{livramento@theor.jinr.ru}

\vspace{.3 in}
\small

\par \vskip .2in \noindent
$^{\dagger}$Instituto de F\'\i sica de S\~ao Carlos; IFSC/USP;\\
Universidade de S\~ao Paulo, USP  \\ 
Caixa Postal 369, CEP 13560-970, S\~ao Carlos-SP, Brazil\\

\par \vskip .2in \noindent
$^{\star}$BLTP, JINR, Dubna 141980, Moscow Region, Russia

\normalsize
\end{center}

\vspace{.5in}

\begin{abstract}

It has been recently proposed a modification of the Skyrme model which admits an exact self-dual sector by the introduction of six scalar fields assembled in a symmetric, positive  and invertible $3\times 3$ matrix $h$.  In this paper we study soft manners of  breaking the self-duality of that model. The crucial observation is that the self-duality equations impose distinct conditions on the three eigenvalues of $h$, and on the three fields lying in the orthogonal matrix that diagonalizes $h$. We keep the self-duality equations for the latter, and break those equations associated to the eigenvalues. We perform the breaking by the addition of kinetic and potential terms for the $h$-fields, and construct numerical solutions using the gradient flow method to minimize the static energy. It is also shown that the addition of just a potential term proportional to the determinant of $h$, leads to a model with an exact self-dual sector, and with self-duality equations differing from the original ones by  just an additional coupling constant.

\end{abstract} 
\end{titlepage}

\section{Introduction}
\label{sec:intro}
\setcounter{equation}{0}

Self-duality plays a prominent role in many areas of Physics, from condensed matter to  high energy physics and  cosmology. The key ingredient for the appearance of self-dual sectors in a given theory is the existence of a (homotopic) topological charge  that admits an integral representation, i.e. there is a density of topological charge \cite{genbps}. The invariance of that charge under any smooth variation of the fields leads, through the integral representation, to local identities which are in general second order partial differential equations satisfied by any smooth field configuration. Together with the self-duality equations, which are first order partial differential equations, those identities imply the second order (dynamical) Euler-Lagrange equations of a given field theory. In addition, in many cases the self-duality leads to a lower bound on the static energy (or Euclidean action) determined by the topological charge and that is saturated by the self-dual solutions. Therefore, on each topological sector such self-dual solutions have the minimum allowed energy and so they are very stable. 

In this paper we want to study the partial breaking of the self-duality, and try to explore the consequences it has on the physics of the remaining quasi-self-dual sector. We shall do that in the context of a Skyrme model in $(3+1)$ dimensions. As it is well known the original Skyrme model \cite{skyrme1,skyrme2} does not possess an exact non-trivial self-dual sector \cite{mantonruback}.  Several modifications of the Skyrme model have been proposed to accommodate a self-dual sector \cite{adam1,adam2,adam_prl,sut_tower,sut_naya_1,sut_naya_2,bpswojtek,bpsshnir}. We shall consider in this paper the  model proposed in \cite{laf2017} defined, in $(3+1)$-dimensional Minkowski space-time, by the action  
\be
S_1= \int d^4x\left[ \frac{m_0^2}{2}\, h_{ab}\,R^a_{\mu}\,R^{b\,,\, \mu}-\frac{1}{4\,e_0^2}\, h^{-1}_{ab}\,H^a_{\mu\nu}\,H^{b\,,\,\mu\nu}\right]
\lab{modelintro}
\ee
where, like in the usual Skyrme model, $R^a_{\mu}$ are the components of the Maurer-Cartan form, i.e. $i\,\partial_{\mu}U\,U^{\dagger}\equiv R^a_{\mu}\,T_a$, with $U$ being a group element of $SU(2)$, and $T_a$ being a basis of its Lie algebra, satisfying
\be
\sbr{T_a}{T_b}=i\,\ve_{abc}\,T_c \; ;\; \qquad\qquad\qquad\qquad 
{\rm Tr}\(T_a\,T_b\)=\kappa\, \delta_{ab}
\lab{su2killing}
\ee
with $\kappa$  being a constant depending upon the representation ($\kappa=1/2$ for the spinor representation, and $\kappa=2$ for the triplet (adjoint) representation). $H^a_{\mu\nu}$ is the curl of that form, i.e. $H^a_{\mu\nu}\equiv \partial_{\mu} R^a_{\nu}-\partial_{\nu} R^a_{\mu}$, and $m_0$ and $e_0$ are coupling constants, of dimension of mass and dimensionless respectively. The model possesses, in addition to the three chiral fields (pions) parameterizing $U$, six extra scalar fields assembled in the symmetric and invertible matrix $h_{ab}$, $a,b=1,2,3$. 
For the static energy associated to \rf{modelintro} to be positive it is required that the eigenvalues of matrix $h$ must also  be positive. 

The properties of such a model have been studied in great detail in \cite{us}, and in \cite{usfalse} a modification of it has been applied to nuclear matter. By coupling it to a fluid theory, where the order parameter is a fractional power of the density of baryonic charge, it was possible to reproduces the bulk behaviour of the binding energy and the radii of 265 nuclei. Such list of nuclei contains all the stable nuclei up to $^{208}$Pb, and above that,  nuclei with a half-life greater than $10^3$ years, up to $^{240}$Pu. The values of such quantities are reproduced with an excellent accuracy (about $1\%$ for both the radius and the binding energy) for the quasi-stable nuclei with mass number equal to $20$ or greater \cite{usfalse}. The error increases for light nuclei with mass number below $20$. The main properties of the model \rf{modelintro}, studied in \cite{us},  can be summarized as follows:

{\em i)} The self-dual sector is defined by the nine self-duality equations 
\be
\lambda\,h_{ab}\,R^b_i= \frac{1}{2}\,\ve_{ijk}\,H^a_{jk}\;;\qquad\qquad\qquad \qquad\lambda\equiv \pm m_0\,e_0 
\lab{selfdualeqs}
\ee
where the indices $a,b=1,2,3$, refer to the group indices, and  $i,j,k=1,2,3$, to the space coordinates $x_i$. The self-duality equations \rf{selfdualeqs} imply  the nine static Euler-Lagrange equations, three of them associated to the  $U$-fields, and also the six equations associated to the scalar fields assembled in $h_{ab}$. 

{\em ii)} The self-dual sector, defined by \rf{selfdualeqs}, and the static sector of the theory \rf{modelintro} are equivalent, i.e. any static solution is self-dual and vice-versa. 

{\em iii)} The static Euler-Lagrange equations associated to the scalar fields $h_{ab}$ imply the self-duality equations \rf{selfdualeqs}, and so indirectly imply also the static Euler-Lagrange equations associated to the $U$-fields. 

{\em iv)} The introduction of the six scalars $h_{ab}$ makes the static sector of \rf{modelintro} invariant under conformal transformations in $\IR^3$, i.e. the self-duality equations \rf{selfdualeqs}, the nine static Euler-Lagrange equations associated to \rf{modelintro}, as well as the static energy
\be
E_1= \int d^3x\left[ \frac{m_0^2}{2}\, h_{ab}\,R^a_{i}\,R^{b}_{i}+\frac{1}{4\,e_0^2}\, h^{-1}_{ab}\,H^a_{ij}\,H^{b}_{ij}\right]
\lab{staticenergy1}
\ee
are all invariant under the conformal group $SO(3,2)$. The infinitesimal conformal transformations in $\IR^3$ are given by $\delta x_i=\zeta_i$ with $\partial_i\zeta_j+\partial_j\zeta_i=2\,D\,\delta_{ij}$, with $D$ vanishing for translations and rotations, it is constant for dilatations, and it is linear in the $x_i$'s for the special conformal transformations. The $U$ fields are scalars under the conformal group, i.e. $\delta U=0$, and the $h$-fields have conformal weight $-1$, i.e.  $\delta h_{ab}=-D\,h_{ab}$. 

{\em v)} The self-duality leads to a lower bound on the static energy \rf{staticenergy1}, and for the self-dual solutions such a bound is saturated as
\be
E_1^{\rm BPS}= 48\,\pi^2\frac{\mid m_0\mid}{\mid e_0\mid}\, \mid Q\mid
\lab{staticenergy1bps}
\ee
where $Q$ is the topological charge 
\be
 Q= \frac{i}{48\,\pi^2}\int d^3x\; \ve_{ijk}\,\trace\(R_i\,R_j\,R_k\)
 \lab{topcharge}
 \ee
 which gives the winding number of the maps $S^3\rightarrow SU(2)$, where $S^3$ is  $\IR^3$ with the spatial infinity identified to a point. Remember that in order to have finite energy solutions the $U$-field must go to a constant at infinity and so, for topological considerations, one can consider such an identification. In \rf{topcharge} we have used the normalized trace
 \be
\trace\(T_a\,T_b\)=\frac{1}{\kappa}\,{\rm Tr}\(T_a\,T_b\)=\delta_{ab}
\lab{normtrace}
\ee
In additional, the sign of $Q$ and $\lambda$ in \rf{selfdualeqs} are related through
\be {\rm sign}\(Q\,\lambda\)=-1 \lab{oposites}\ee

{\em vi)} An important role is played by the real and symmetric matrix
\be
\tau_{ab}\equiv R_i^a\,R_i^b
\lab{taudef}
\ee
If ${\rm det}\,\tau=0$, then the only possible static solution is $U={\rm constant}$. If ${\rm det}\,\tau\neq 0$, then the self-duality equations \rf{selfdualeqs} imply that the matrix $h$ is determined from the $U$-fields configuration by
\be
h_{\rm BPS}=\frac{\sqrt{{\rm det}\,\tau}}{\mid m_0\,e_0\mid}\; \tau^{-1}
\lab{hbps}
\ee
That means that the self-duality equations are satisfied by any non-trivial configuration of the $U$-fields, and the $h$-fields adjust themselves to solve the self-duality, taking the form \rf{hbps}. For the BPS field configurations \rf{hbps} the quadratic and quartic terms in the space-time derivatives of \rf{modelintro} give exactly the same contribution to the total energy \rf{staticenergy1}, i.e.
\be
\frac{m_0^2}{2}\,\int d^3x \, \(h_{{\rm BPS}}\)_{ab}\,R^a_{i}\,R^{b}_{i}=\frac{1}{4\,e_0^2}\int \, d^3x \, \(h_{{\rm BPS}}\)^{-1}_{ab}\,H^a_{ij}\,H^{b}_{ij}=24\,\pi^2\frac{\mid m_0\mid}{\mid e_0\mid}\, \mid Q\mid
\lab{relations1}
\ee
and the topological charge \rf{topcharge} can be written in terms of the eigenvalues of $h_{{\rm BPS}}$ as 
\be Q = -\frac{\lambda^3}{16\,\pi^2}\,\int d^3x \, \det h_{{\rm BPS}} \lab{relations2}\ee

{\em vii)} As the matrices $h$ and $\tau$ are and symmetric, they can be diagonalized by orthogonal transformations, i.e. 
\br
h&=&M\,h_D\,M^T\;;\qquad\qquad M\,M^T=\one\;;\qquad\qquad \(h_D\)_{ab}=\vp_a\,\delta_{ab}
\nonumber\\
\tau&=&N\,\tau_D\,N^T\;;\qquad\qquad\;\; N\,N^T=\one\;;\qquad\qquad\;  \(\tau_D\)_{ab}=\omega_a\,\delta_{ab}
\lab{diagonalhtau}
\er
When the self-duality equations \rf{selfdualeqs} hold true, and so \rf{hbps} is valid, we have that the matrices $h$ and $\tau$ commute and so can be diagonilized simultaneously, i.e. $M=N$, and the eigenvalues are related by
\be 
\vp_a = \frac{1}{\mid m_0\,e_0 \mid}\,\sum_{b,\,c=1}^3\frac{\mid\varepsilon_{abc}\mid}{2}\,\sqrt{\frac{\omega_b\,\omega_c}{\omega_a}}
\lab{useful}
\ee
or equivalently
 \be
 \frac{\omega_1}{\vp_2\,\vp_3}=\frac{\omega_2}{\vp_1\,\vp_3}=\frac{\omega_3}{\vp_1\,\vp_2}=m_0^2\,e_0^2
 \lab{eigenvaluerel}
 \ee
and so
\br
 \frac{\omega_1\,\omega_2}{\vp_3}=\frac{\omega_1\,\omega_3}{\vp_2}=\frac{\omega_2\,\omega_3}{\vp_1}&=&m_0^4\,e_0^4\,\vp_1\,\vp_2\,\vp_3
 \lab{eigenvaluerel2}\\
 \omega_1\,\vp_1=\omega_2\,\vp_2=\omega_3\,\vp_3&=&m_0^2\,e_0^2\,\vp_1\,\vp_2\,\vp_3
 \nonumber
 \er
{\em viii)} The action \rf{modelintro} is invariant under the global symmetry $SU(2)_L\otimes SU(2)_R$ defined by the transformations 
\be
U\rightarrow g_L\,U\;;\qquad\qquad R_{\mu}^a\rightarrow d_{ab}\(g_L\)\,R_{\mu}^b \;;\qquad\qquad 
h_{ab}\rightarrow d_{ac}\(g_L\)\, h_{cd}\,d_{db}^T\(g_L\)
\lab{leftsymm}
\ee
and
\be
U\rightarrow U\,g_R\;;\qquad\qquad\qquad R_{\mu}^a\rightarrow R_{\mu}^a \;;\qquad\qquad \qquad
h_{ab}\rightarrow h_{ab}
\lab{rightsymm} 
\ee
with $g_{L/R}\in SU(2)_{L/R}$, and where $d\(g\)$ is the $3\times 3$ matrix for the group element $g$ in the adjoint (triplet) representation of $SU(2)$, i.e. 
\be
g\,T_a\, g^{-1}=T_b\,d_{ba}\(g\)\;;\qquad\qquad\qquad d\(g_1\)\,d\(g_2\)=d\(g_1\,g_2\)
\lab{adjoint}
\ee

 In the self-dual model described above the $h$-fields are not propagating, as they enter into the action \rf{modelintro} through the coupling to the $U$-fields by contracting the group indices, and there is not kinetic term for them. In this paper we want to break the self-duality by adding kinetic and potential terms for the $h$-fields. Note that the $3\times 3$ orthogonal matrix $M$, diagonalizing $h$ in \rf{diagonalhtau}, has three independent components. Therefore, we can take the six scalar fields assembled in the matrix $h$ to be those three components of $M$ and the three eigenvalues $\vp_a$, $a=1,2,3$, introduced in \rf{diagonalhtau}. The eigenvalues of $h$ are invariant under the transformations $SU(2)_L\otimes SU(2)_R$, given in \rf{leftsymm} and \rf{rightsymm}, and the $M$-fields are invariant under \rf{rightsymm}, and transform as $M\rightarrow d\(g_L\)\,M$ under \rf{leftsymm}. In addition, the $M$-fields are scalars under the conformal group $SO\(3\,,\,2\)$, and the $\vp$-fields have conformal weight $-1$, i.e. $\delta \vp_a= -D\,\vp_a$ (see item {\em iv)} above). With such a decomposition of fields, a kinetic term for the $h$-fields takes the form ${\rm Tr}\(\partial_{\mu}h\)^2=\(\partial_{\mu}\vp_a\)^2+{\rm Tr}\(\sbr{M^T\partial_{\mu}M}{h_D}\)^2$. Since $\vp_a$ and $M^T\partial_{\mu}M$ are invariant under the chiral transformations $SU(2)_L\otimes SU(2)_R$, given in \rf{leftsymm} and \rf{rightsymm}, we shall take an arbitrary linear combination of those terms. So, we shall consider the theory 
 \be 
 S = S_1 + S_2 
 \lab{full}
 \ee
 with $S_1$ given in  \rf{modelintro}  and  
\be
S_2=\int d^4x\left[
 \frac{\mu_0^2}{2}\,\left[ \sum_{a=1}^3 \kappa_a \(\partial_{\mu}\vp_a\)^2+ \kappa_4\, {\rm Tr}\(\sbr{M^T\partial_{\mu}M}{h_D}\)^2\right]- {\cal V} \(\vp\)
 -\frac{\beta_3^2}{2}\, {\rm Tr}\(\one - U\)\right] 
 \lab{s2def}
 \ee
where $\m_0$ and $\beta_3$,  are coupling constants of dimension of mass and of mass$^2$ respectively, and $\kappa_a$, $a=1,2,3,4$ are dimensionless parameters. The $\beta_3$-term is a mass term for the (pion) $U$-fields. 

If $\beta_3=0$, the action \rf{s2def} is  invariant under \rf{leftsymm} and \rf{rightsymm}. However, if $\beta_3\neq 0$, the symmetry $SU(2)_L\otimes SU(2)_R$ is broken to the diagonal subgroup $g_R=g_L^{-1}\equiv g^{-1}$, i.e. $U\rightarrow g\,U\,g^{-1}$. The conformal symmetry of the static energy associated to \rf{s2def}, i.e. 
\be
E_2=\int d^3x\left[
 \frac{\mu_0^2}{2}\,\left[ \sum_{a=1}^3 \kappa_a \(\partial_{i}\vp_a\)^2+ \kappa_4\, {\rm Tr}\(\sbr{M^T\partial_{i}M}{h_D}\)^2\right]+ {\cal V} \(\vp\)+\frac{\beta_3^2}{2}\, {\rm Tr}\(\one - U\)\right] 
 \lab{e2def}
 \ee
is  broken by the $\beta_3$-term, as $U$ is a scalar under $SO(3,2)$, and so it does not compensate the transformation of the volume $d^3x$. The kinetic, and possible mass terms in ${\cal V}$, for the $\vp_a$-fields  also break the conformal symmetry, as the $\vp_a$-fields have conformal weight $-1$. The potential  ${\cal V}$ does not break the conformal symmetry only if it is cubic in the $\vp_a$-fields.

In this paper we shall break the self-duality only partially, as we shall impose that the matrices $h$ and $\tau$ still commute (see \rf{diagonalhtau})
\be
\sbr{h}{\tau}=0\qquad\qquad \leftrightarrow \qquad\qquad M=N
\lab{halfbpscond}
\ee
 Since \rf{hbps} ceases to be true, we have that the eigenvalues of $h$ and $\tau$ will not be related  by \rf{useful}, \rf{eigenvaluerel} and  \rf{eigenvaluerel2} anymore. 

 From \rf{taudef} we see that the entries of the matrix $\tau$ are functionals of the $U$-fields and their first derivatives. Consequently, the entries of the orthogonal matrix $N$ and the eigenvalues $\omega_a$, introduced in \rf{diagonalhtau}, are also functional of $U$-fields and their first derivatives.  Therefore, the condition \rf{halfbpscond} is saying that the three $M$-fields are determined from the $U$-fields, in a way similar to that in \rf{hbps}, where $h$ is determined from the $U$-fields, when the full self-duality equations are valid. For that reason we can consider \rf{halfbpscond} as quasi-self-duality equations. As we explain in the section \ref{sec:quasisd} such a condition introduces nice simplifications in the model. 

There are some particular interesting  cases of the full theory \rf{full} that break the self-duality in a soft manner, as we discuss in section \ref{sec:first}. In addition to the three quasi-self-duality equations \rf{halfbpscond}, one can impose algebraic relations among the eigenvalues of the matrices $h$ and $\tau$, such that the variation of the energy functional $E_1$ with respect to the $U$-fields, becomes proportional to the variation of the topological charge, and so vanishes identically. In other words, algebraic relations among the eigenvalues of $h$ and $\tau$, solve the part of the Euler-Lagrange equations, associated to the $U$-fields, coming from $E_1$. A further consequence of such algebraic relations is that  the matrix $h$ becomes proportional to self-dual matrix $h_{BPS}$, given in \rf{hbps}. It then follows that  the variation of density of the energy functional $E_1$ with respect to the $\vp$-fields becomes proportional to the variation of ${\rm det}\,h$, with respect to the same fields. Therefore, we can solve the Euler-Lagrange equations associated to the $\vp$-fields by restricting the energy functional $E_2$ to be proportional  to ${\rm det}\,h$, i.e. by restricting \rf{e2def} to the case where $\kappa_a=0$, $a=1,2,3,4$, $\beta_3=0$, and ${\cal V}=\frac{\beta_{{\cal V}}^2}{2}\, {\rm det}\, h=\frac{\beta_{{\cal V}}^2}{2}\, \vp_1\,\vp_2\,\vp_3 $, with $\beta_{{\cal V}}$  a real dimensionless coupling constant.  So, the static energy of such model becomes 
  \be
E_{\rm quasi-sd}= \int d^3x\left[ \frac{m_0^2}{2}\, h_{ab}\,R^a_{i}\,R^{b}_{i}+\frac{1}{4\,e_0^2}\, h^{-1}_{ab}\,H^a_{ij}\,H^{b}_{ij}+ \frac{\beta_{{\cal V}}^2}{2}\, {\rm det}\, h\right]
\lab{firstredu}
\ee
The model \rf{firstredu} has an exact self-dual sector where the self-duality equation which differs from \rf{selfdualeqs} by a multiplicative parameter, i.e. it is given by 
\be
\frac{\lambda}{\alpha}\,h_{ab}\,R^b_i= \frac{1}{2}\,\ve_{ijk}\,H^a_{jk}\;;\qquad\qquad\qquad \qquad\lambda\equiv \pm m_0\,e_0 
\lab{selfdualeqsgene}
\ee
where $\alpha$ is a monotonically decreasing function of the strength $\beta_{{\cal V}}$ of the potential, with $\alpha=1$ for $\beta_{{\cal V}}=0$ (the details are given in section \ref{sec:first}). 
 
Note that $\int d^3x\,{\rm det}\,h$, is invariant under conformal transformations in $\IR^3$. Therefore, such soft manner of breaking the self-duality preserves all the symmetries of the self-dual Skyrme model \rf{modelintro}, namely the global symmetries $SU(2)_L\otimes SU(2)_R$, defined by the transformations \rf{leftsymm} and \rf{rightsymm}, as well as the conformal symmetry in the three dimensional space. Note in addition that, since $h=\alpha\,h_{{\rm BPS}}$, with $h_{{\rm BPS}}$ given in \rf{hbps}, the $h$-fields still act as spectators of the $U$-fields, which in turn remain totally free. Therefore, such a theory also leads to an infinite number of exact topological solutions for any value of $Q$ and extends the results obtained in \cite{us}. The total energy $E=E_1+E_2$ is proportional to $\mid Q\mid$, but the proportionality constant is a monotonic increasing function of $\beta_{{\cal V}}$, i.e. the strength of the potential ${\cal V}$.  

In order to construct solutions  for the full theory \rf{full}, subjected to the quasi-self-duality equations \rf{halfbpscond}, we shall work with the  so-called rational map ansatz \cite{mantonbook,rational1, rational2} for the $U$-fields, which is described  in section \ref{sec:secondholomorphic}. In such an ansatz, spheres of radius $r$, in the spatial submanifold $\IR^3$, are stereographically projected on a plane parametrized by a complex coordinate $w$. The $U$-fields are then given by a profile function $f$, depending only on the radial distance $r$, and a complex field $u$ which is an holomorphic function of  $w$ and a map between two-spheres. In such an ansatz, the first two eigenvalues of the matrix $\tau$,  defined in \rf{taudef}, become equal, and the third one is a function of the radial variable only, i.e. $\omega_1=\omega_2$ and $\omega_3=\omega_3\(r\)$. 

As a consequence of the quasi-self-duality equations \rf{halfbpscond}, the Euler-Lagrange equations for the $M$-fields become differential equations to be satisfied by the $U$-fields, in addition to their own Euler-Lagrange equations. That would be a too restrictive condition on the $U$-fields. However, we observe that by imposing that the eigenvalues  of the matrix $h$,  depend only on the radial distance $r$, i.e. $\vp_a=\vp_a\(r\)$, and in addition that the first two eigenvalues are equal, i.e. $\vp_1=\vp_2$, we solve the Euler-Lagrange equations for the $M$-fields automatically.  Under such conditions the Euler-Lagrange equations of $u$ and ${\bar u}$ fields are also automatically satisfied. 

The drawback of that procedure is that only some special configurations of the $u$-field with unity topological degree  can be solutions of our quasi-self-dual model, inside the holomorphic ansatz. Such fixing of the $u$-field imposes radial symmetry to the topological charge density and restricts the construction of topological solutions with large values of $Q$ by choosing properly the boundary conditions of the profile $f$ function, which may lead to unstable static solutions for $\mid Q\mid \,\geq 2$.  However, the advantage of the procedure  is that we are left to solve  only three ordinary differential equations, which  correspond to the Euler-Lagrange equations for the profile function $f\(r\)$, and for the $\vp_1\(r\)$ (equal to $\vp_2$) and $\vp_3\(r\)$ fields. Those equations are solved numerically using the gradient flow method to minimize the static energy of the system. 

Such an analysis of the static sector of the full theory \rf{full} is very important to study the effect of allowing the scalar fields in the matrix $h$, to be propagating fields.  From the results of section \ref{sec:numerics} one observes, as the strength of the kinetic and potential terms for the $h$-fields increase,  the eigenvalues $\vp_a$ of the matrix $h$ tend to grow at the origin and to fall exponentially faster at large distances. 

The paper is organized as follows. In the section \ref{sec:quasisd} we obtain all the  nine Euler-Lagrange equations for the static version of \rf{full} inside the quasi-self-dual ansatz \rf{halfbpscond}. In  section \ref{sec:first} we show how some special algebraic relations among the eigenvalues of the matrices $h$ and $\tau$ lead to an exact self-dual sector of the model \rf{firstredu}. In section \ref{sec:second} we consider the static version of full theory \rf{full}, and in subsection \ref{sec:secondholomorphic} we construct an holomorphic ansatz for it, compatible with \rf{halfbpscond}.  In subsection \ref{sec:elequationsholomorphic} we analyse the Euler-Lagrange equations of \rf{full} within the holomorphic and \rf{halfbpscond} ansatz\"e. The numerical solutions of those equations is constructed in subsection \ref{sec:numerics} for a quadratic potential for the $h$-fields. Our conclusions are presented in section \ref{sec:conclusion}. The Appendix \ref{sec:algebraicsol} presents the proof of the algebraic relations used in section \ref{sec:first},  and Appendix \ref{app:proof} shows  why only the solutions with unity baryonic charge satisfy the conditions of section \ref{sec:elequationsholomorphic}. Appendix \ref{app:numerics} presents some details of our numerical methods.

\section{The Quasi-Self-Duality}
\label{sec:quasisd}
\setcounter{equation}{0}

The static energy $E_1$, defined in \rf{staticenergy1}, can be written as 
\be
E_1= \int d^3x\left[ \frac{m_0^2}{2}\, {\rm Tr}\(h\,\tau\)+\frac{1}{4\,e_0^2}\, {\rm Tr}\(h^{-1}\,\sigma\)\right]
\lab{staticenergy1sigma}
\ee
where $\tau$ is defined in \rf{taudef}, and where we have introduced the matrix
\be
\sigma_{ab}\equiv H_{ij}^a\,H_{ij}^b \lab{sigma0}
\ee
The quantities $R_{i} \equiv i\,\partial_{i}U\,U^{\dagger}\equiv R^a_{i}\,T_a$ satisfy the Maurer-Cartan equation
$\partial_{i}R_{j}-\partial_{j}R_{i}+i\,\sbr{R_{i}}{R_{j}}=0$, and so we  have that
\be
H^a_{ij}\equiv \partial_{i} R^a_{j}-\partial_{j} R^a_{i}
=-i\, \trace\(\sbr{R_{i}}{R_{j}}\,T_a\)=\ve_{abc}\,R_{i}^b\,R_{j}^c
\lab{hdef}
\ee
Conjugating both sides of the commutation relations \rf{su2killing} with an $SU(2)$ group element $g$, and using \rf{adjoint}, one gets that
\be
\ve_{abc}\,d_{dc}\(g\)=\ve_{def}\,d_{ea}\(g\)\,d_{fb}\(g\)
\lab{invadjointpre}
\ee
The adjoint representation of $SU(2)$ is a real and unitary representation, and so the matrices $d\(g\)$ are orthogonal. In fact, any orthogonal matrix with determinant $1$ $(-1)$ can be identified with a given matrix $d\(g\)$ $\(-d\(g\)\)$ for some $g\in SU(2)$. Therefore, the orthogonal matrices $M$ and $N$ (as well as their transposes) satisfy \rf{invadjointpre} with a sign given by $\det M=\pm 1$, i.e. 
\be
\ve_{abc}\,M_{dc}\(g\)=\det M\,\ve_{def}\,M_{ea}\(g\)\,M_{fb}\(g\)
\lab{invadjoint}
\ee
and a similar relation for $N$. So, using that fact, \rf{hdef},  and \rf{diagonalhtau} we get that
\be
\sigma_{ab}=  \ve_{acd}\,\ve_{bef}\,\tau_{ce}\,\tau_{df}=\ve_{cde}\,\ve_{cdf}\,\omega_c\,\omega_d\, N_{ae}\,N^T_{fb}=\(N\,\sigma_D\,N^T\)_{ab}
\lab{htaurel}
\ee
where we have defined the matrix 
\be
\(\sigma_D\)_{ab}=\sum_{c,d=1}^3 \ve_{cda}\,\ve_{cdb}\,\omega_c\,\omega_d
\ee
which is diagonal
\be
\sigma_D=2\;\,{\rm diag.}\( \omega_2\,\omega_3\,,\,\omega_1\,\omega_3\,,\, \omega_1\,\omega_2\) \lab{sigma1} 
\lab{sigmadomegarel}
\ee
So, $\sigma$ is diagonalized by the same orthogonal matrix $N$, as $\tau$, and as a consequence of the condition \rf{halfbpscond} we have that
\be
\sbr{h}{\tau}=\sbr{h}{\sigma}=\sbr{\tau}{\sigma}=0
\lab{allcommute}
\ee
Therefore, when considering variations w.r.t. the $M$-fields we have that 
\be \delta^{(M)} h= \sbr{\delta^{(M)} M\, M^T}{h};\qquad\qquad {\rm and}\qquad\qquad\delta^{(M)} h^{-1}= \sbr{\delta^{(M)} M\, M^T}{h^{-1}}\ee and so
\be
\delta^{(M)} E_1= \int d^3x\;{\rm Tr}\left[ \delta^{(M)} M\, M^T\(\frac{m_0^2}{2}\, \sbr{h}{\tau}+\frac{1}{4\,e_0^2}\, \sbr{h^{-1}}{\sigma}\)\right]=0 \lab{Mvariation}
\ee
We then conclude that the Euler-Lagrange equations associated to the $M$-fields, coming from $E_1$, are automatically satisfied, due to the condition  \rf{halfbpscond}, which leads to \rf{allcommute}. Therefore, the non-trivial Euler-Lagrange equations associated to the $M$-fields come from the $\kappa_4$-term in $E_2$, defined in \rf{e2def}, and it is given by 
\be
\partial_i\sbr{h_D}{\sbr{h_D}{M^T\partial_iM}}+\sbr{M^T\partial_iM}{\sbr{h_D}{\sbr{h_D}{M^T\partial_iM}}}=0
\lab{eqforM}
\ee

In addition, using \rf{staticenergy1sigma}, \rf{sigmadomegarel} and the fact that we are assuming that $M=N$, we get that the variation of $E_1$ w.r.t. $\vp_a$ is given by
\be
\delta^{(\vp_a)}\,E_1=\frac{1}{2\,e_0^2}\,\int d^3x\left[ m_0^2\,e_0^2\,\omega_a\,-\frac{1}{2}\,\sum_{b,c=1}^3\mid \ve_{abc}\mid \frac{\omega_b\,\omega_c}{\vp_a^2}\right]\,\delta\vp_a
\lab{varye1vpa}
\ee

Now, when considering variations w.r.t. the $U$-fields we have that $\delta^{(U)}\tau=\sbr{\delta^{(U)} N\, N^T}{\tau}+N\,\delta^{(U)}\tau_D\,N^T$, and a similar relation for  $\sigma$. Therefore, using \rf{halfbpscond}, \rf{sigmadomegarel} and \rf{allcommute}, we get 
\br
\delta^{(U)} E_1&=& \int d^3x\left[ \frac{m_0^2}{2}\, {\rm Tr}\(h_D\,\delta^{(U)}\tau_D \)+\frac{1}{4\,e_0^2}\, {\rm Tr}\(h^{-1}_D\,\delta^{(U)}\sigma_D\)\right]
\nonumber\\
&=&\frac{1}{2\,e_0^2}\, \int d^3x\left[ m_0^2\,e_0^2\, \sum_{a=1}^3\vp_a\,\delta^{(U)}\omega_a + \frac{1}{2}\sum_{a,b,c=1}^3 \mid\ve_{abc}\mid \frac{\delta^{(U)}\(\omega_b\,\omega_c\)}{\vp_a}\right]
\lab{deltaue1}
\er
In addition, we have, from \rf{e2def}, that
\be
\delta^{(U)}E_2=-\frac{\beta_3^2}{2}\,\int d^3x \;\delta^{(U)} {\rm Tr}\(U\)
\lab{deltaue2}
\ee
Consequently, the Euler-Lagrange equations associated to the $U$-fields, coming from $E_1$ and $E_2$, do not involve the $M$-fields. 

Since $M$ is a $3\times 3$ orthogonal matrix, it follows that $M^T\partial_iM$ is a matrix in the adjoint representation of the $SU(2)$ Lie algebra. So, we can write  $M^T\partial_iM=i\,{\cal M}_i^a\,d\(T_a\)$, with $d_{ab}\(T_c\)=i\,\ve_{acb}$. In addition one can show that 
\be
\sbr{h_D}{\sbr{h_D}{d\(T_a\)}}=\frac{1}{2}\sum_{b,c=1}^3 \mid \ve_{abc}\mid \(\vp_b-\vp_c\)^2\, d\(T_a\)
\lab{doublecomhd}
\ee
Therefore
\br
{\rm Tr}\(\sbr{M^T\partial_{i}M}{h_D}\)^2&=&-{\rm Tr}\(M^T\partial_{i}M\sbr{h_D}{\sbr{h_D}{M^T\partial_{i}M}}\)
\\
&=&\sum_{a,b,c=1}^3 \mid \ve_{abc}\mid \(\vp_a-\vp_b\)^2\,{\cal M}_i^c\,{\cal M}_i^c
\er
where we have used the fact that ${\rm Tr}\(d\(T_a\)\,d\(T_b\)\)=2\,\delta_{ab}$. Consequently, the variation of $E_2$, given in \rf{e2def}, w.r.t. $\vp_a$ is 
\be
\delta^{(\vp_a)}\,E_2=\int d^3x\left[
 \mu_0^2\,\left[ - \kappa_a \partial_{i}^2\vp_a+ \kappa_4\, \sum_{b,c=1}^3 \mid \ve_{abc}\mid \(\vp_a-\vp_b\)\,{\cal M}_i^c\,{\cal M}_i^c\right]
+ \frac{\delta\,{\cal V}}{\delta \vp_a} \right] \delta \vp_a
 \lab{varye2vpa}
 \ee
 
 In some of our applications it will be useful to treat the quantities $R_i^a$ as a $3\times 3$ matrix with the following ordering of rows and columns $R_i^a=\(R\)_{ia}$, $i=1,2,3$, and $a=1,2,3$. Therefore
 \be
 \ve_{ijk}\,R_i^a\,R_j^b\,R_k^c=\ve_{abc}\,\ve_{ijk}\,R_{i1}\,R_{j2}\,R_{k3}=\ve_{abc}\,{\rm det}\,R
 \lab{detrrel}
 \ee
 Then from \rf{normtrace} we have that
 \be
 \ve_{ijk}{\widehat {\rm Tr}}\(R_i\,R_j\,R_k\)=i\,3\,{\rm det}\,R
 \ee
 But from \rf{taudef} we have that ${\rm det}\,\tau=\({\rm det}\,R\)^2$. Therefore, the topological charge \rf{topcharge} can be written as 
 \be
 Q=- \frac{\ve}{16\,\pi^2}\,\int d^3x\, \sqrt{{\rm det}\,\tau};\qquad \quad{\rm det}\,R=\ve  \sqrt{{\rm det}\,\tau};\qquad\quad \ve=\pm 1
 \lab{topchargedettau}
 \ee
  Note that the eigenvalues of the matrix $\tau$, given in \rf{taudef}, are all non-negative since if $v_a$ is an arbitrary real vector then 
 \be
 v^T\,\tau\,v=\sum_{i=1}^3\(v_a\,R_i^a\)^2\geq 0
 \lab{positivetaueigen}
 \ee
 The topological charge is invariant under any (homotopic) smooth variation of the $U$-fields, i.e. $ \delta^{(U)} Q=0$, and consequently  the eigenvalues of $\tau$ have to satisfy
 \be
  \int d^3x\,  \delta^{(U)}\sqrt{\omega_1\,\omega_2\,\omega_3}=0
 \lab{nicereltop}
 \ee
 Such a relation will be very useful, in section \ref{sec:first}, in the construction of models that break the self-duality in a soft manner. 
 
 \section{The first type of quasi-self-dual model}
\label{sec:first}
\setcounter{equation}{0}

By considering the coefficient of $\delta^{(U)}\omega_a$ in \rf{deltaue1}, for each value of $a=1,2,3$, we observe that if we impose
 \br
 m_0^2\,e_0^2\,\vp_1+\frac{\omega_3}{\vp_2}+\frac{\omega_2}{\vp_3}&=& \Lambda\,\sqrt{\frac{\omega_2\,\omega_3}{\omega_1}}
 \nonumber\\
  m_0^2\,e_0^2\,\vp_2+\frac{\omega_3}{\vp_1}+\frac{\omega_1}{\vp_3}&=& \Lambda\,\sqrt{\frac{\omega_1\,\omega_3}{\omega_2}}
 \lab{funnyrel}\\
  m_0^2\,e_0^2\,\vp_3+\frac{\omega_2}{\vp_1}+\frac{\omega_1}{\vp_2}&=& \Lambda\,\sqrt{\frac{\omega_1\,\omega_2}{\omega_3}}
 \nonumber
 \er
 with $\Lambda$ being an arbitrary constant with dimension of mass, which is non-negative since the eigenvalues $\vp_a$ and $\omega_a$, $a=1,2,3$, are non-negative, then, as a consequence of \rf{nicereltop}, \rf{deltaue1} becomes
 \be
 \delta^{(U)} E_1=\frac{\Lambda}{e_0^2}\, \int d^3x\, \delta^{(U)}\sqrt{\omega_1\,\omega_2\,\omega_3}=0
 \lab{vareiequalvartopcharge}
 \ee
  In other words, the algebraic relations \rf{funnyrel} imply that the part of Euler-Lagrange equations associated to the $U$-fields, coming from  $E_1$, are satisfied. Therefore, if we drop the pion mass term from \rf{e2def}, i.e. take $\beta_3=0$, we are left to consider only the Euler-Lagrange equations associated to the $h$-fields.  So, as far as the $U$-fields are concerned, the algebraic relations \rf{funnyrel} play the same role, in the theory \rf{full}, as the (differential) self-duality equations \rf{selfdualeqs}  in the self-dual Skyrme model \rf{modelintro}.

 The solutions of the algebraic equations \rf{funnyrel} are constructed in Appendix \ref{sec:algebraicsol}. There are basically three types of solutions, but since we need the eigenvalues of the matrix $h$ to be positive, only one type is adequate for our applications. It is given by 
 
 \be 
 \vp_a = \frac{\alpha}{\mid m_0\,e_0 \mid}\,\sum_{b,\,c=1}^3\frac{\mid\varepsilon_{abc}\mid}{2}\,\sqrt{\frac{\omega_b\,\omega_c}{\omega_a}}
 \lab{useful2}
 \ee
 with $\alpha$ being related to $\Lambda$  by
\be 
\alpha=\frac{1}{2}\(\frac{\Lambda}{m_0\,e_0}\pm\sqrt{\frac{\Lambda^2}{m_0^2\,e_0^2}-8}\) \qquad\qquad {\rm with} \qquad\qquad \Lambda \geq 2\,\sqrt{2}\, \mid m_0\,e_0\mid; \qquad \alpha\geq 0
\ee 
Note that \rf{useful2} differs from \rf{useful} only by the factor $\alpha$, and one can check that \rf{useful2}, together with \rf{halfbpscond},  imply that the matrix $h$ has the form 
\be
h = \alpha\,h_{BPS}; \qquad\qquad {\rm with} \qquad\qquad h_{\rm BPS}=\frac{\sqrt{{\rm det}\,\tau}}{\mid m_0\,e_0\mid}\; \tau^{-1} 
\lab{ansatzh}
\ee
Using the definition of $\tau$ in \rf{taudef},  one gets that \rf{ansatzh} leads to
\be
\mid m_0\,e_0\mid\, h_{ac}\,R_i^c\,R_i^b=\alpha\,\sqrt{{\rm det}\,\tau}\,\delta_{ab}\qquad\rightarrow\qquad
\mid m_0\,e_0\mid\, h_{ac}\,R_i^c=\alpha\,\sqrt{{\rm det}\,\tau}\,\(R^{-1}\)_{ai}
\lab{rinverserel1}
\ee
Using \rf{detrrel} and \rf{topchargedettau}, one gets that
\be
\frac{1}{2}\,\ve_{abc}\,\ve_{ijk}\,R_i^a\,R_j^b=\pm \sqrt{{\rm det}\,\tau}\,\(R^{-1}\)_{ck}
\lab{rinverserel2}
\ee
Combining \rf{rinverserel1} and \rf{rinverserel2} one gets that  the $h$-fields must satisfy a generalized version of the self-dual equations \rf{selfdualeqs} given by 
\be
\lambda\,h_{ab}\,R^b_i= \frac{\alpha}{2}\,\ve_{ijk}\,H^a_{jk}\;;\qquad\qquad\qquad \qquad\lambda\equiv \pm m_0\,e_0 
\lab{selfdualeqsgene}
\ee
where we have used \rf{hdef}. 

Using \rf{useful2} one gets that \rf{varye1vpa} becomes 
\be
\delta^{(\vp_a)}\,E_1=\frac{m_0^2}{2}\,\int d^3x\left[ 1-\frac{1}{\alpha^2}\right]\,\omega_a\,\delta\vp_a=\frac{m_0^4\,e_0^2}{2}\,\int d^3x\left[ 1-\frac{1}{\alpha^2}\right]\,\frac{1}{\alpha^2}\,\frac{\delta\,{\rm det}\,h}{\delta\,\vp_a}\,\delta\vp_a
\ee
where, in the last equality, we have used the fact that \rf{useful2} implies that
\ \be
 \frac{\omega_1}{\vp_2\,\vp_3}=\frac{\omega_2}{\vp_1\,\vp_3}=\frac{\omega_3}{\vp_1\,\vp_2}=\frac{m_0^2\,e_0^2}{\alpha^2}
 \lab{eigenvaluerelalpha}
 \ee
 Therefore, if we choose all the terms in $E_2$, given in \rf{e2def}, to vanish except for the potential term which we take to be proportional to ${\rm det}\,h$, i.e. we assume \rf{firstredu}, we solve the Euler-Lagrange equations associated to the $\vp$-fields. Using the notation of \rf{firstredu} we then get that\footnote{Note that there two solutions for $\alpha$, namely $\alpha^2 = 2/(1\pm\sqrt{1+4\,\vartheta})$. But since $\alpha$ and $\vartheta$ are non-negative parameters, then $\alpha$ is reduced to \rf{alpha}.}
\be
\beta_{{\cal V}}^2=m_0^4\,e_0^2\,\left[ \frac{1}{\alpha^4}-\frac{1}{\alpha^2}\right];\quad {\rm or}\quad \alpha\(\vartheta\) = \sqrt{\frac{2}{1+\sqrt{1+4\,\vartheta}}}\,;\quad {\rm with}\quad\vartheta \equiv \frac{\beta_{{\cal V}}^2}{m_0^4\,e_0^2} 
\lab{alpha}
 \ee
 The Euler-Lagrange equations for the $M$-fields, given in \rf{eqforM}, comes from the $\kappa_4$-term in $E_2$, given in \rf{e2def}. Since we have dropped  that term we do not have such an equation in this model.
 
 Therefore, the solutions of the modified self-duality equations \rf{selfdualeqsgene}, are static solutions of the theory defined by the following static energy functional
 \be
E_{\rm quasi-sd}= \int d^3x\left[ \frac{m_0^2}{2}\, h_{ab}\,R^a_{i}\,R^{b}_{i}+\frac{1}{4\,e_0^2}\, h^{-1}_{ab}\,H^a_{ij}\,H^{b}_{ij}+ \frac{\beta_{{\cal V}}^2}{2}\, {\rm det}\, h\right]
\lab{staticenergyfirstquasisd}
\ee
We have then obtained an extension of the theory \rf{staticenergy1}, by the addition of a potential proportional to ${\rm det}\, h$, which admits an exact self-dual sector. The self-duality equations for the two theories differ just by a multiplicative constant in one of its two terms.  Such particular extension of the BPS theory \rf{staticenergy1} preserves the conformal invariance in three spacial dimensions as well the the global symmetry $SU(2)_L\otimes SU(2)_R$ defined by the transformations \rf{leftsymm} and \rf{rightsymm}.

Using \rf{topchargedettau} one gets that  the static energy \rf{staticenergyfirstquasisd}  evaluated  on the solutions of the self-duality equations  \rf{selfdualeqsgene} becomes 
\be
E_{\rm quasi-sd}= 24\,\pi^2\,\frac{\mid m_0 \mid}{\mid e_0\mid}\,\(\alpha+\frac{1}{\alpha}+\frac{\vartheta}{3}\,\alpha^3\)\,\mid Q \mid
\lab{energyquasi-sd}
\ee
Writing \rf{energyquasi-sd} in terms of the BPS static energy $E_1^{BPS}$, given in \rf{staticenergy1bps}, and using \rf{alpha} we obtain 
\be 
E_{\rm quasi-sd} = E_1^{\rm BPS} \,\frac{\sqrt{2}}{3}\,\frac{2+\sqrt{1+4\,\vartheta}}{\sqrt{1+\sqrt{1+4\,\vartheta}}}\lab{enerdet2}
;\qquad\qquad {\rm with}\qquad\qquad E_1^{\rm BPS}=48\,\pi^2\frac{\mid m_0 \mid}{\mid e_0 \mid}\,\mid Q \mid 
\lab{energyend}
\ee
which is monotonic increasing on $\vartheta$. In addition, on the weak coupling regime $\vartheta \ll 1 $ the static energy \rf{enerdet2} becomes $E\approx E_1^{\rm BPS}\,\(1+\vartheta/6-\vartheta^2/8+\mathcal{O}\(\vartheta^3\)\)$, and on the strong coupling regime $\vartheta \gg 1$ we have ${E\approx E_1^{\rm BPS}\,\(2\,\vartheta^{\frac{1}{4}}/3+\vartheta^{-\frac{1}{4}}/2-\vartheta^{-\frac{3}{4}}/16\right.} {\left.   +\mathcal{O}\(\vartheta^{-5/4}\)\)}$. 

Therefore, the addition of a potential term proportional to ${\rm det}\,h$ to the theory \rf{modelintro}, does not really break the self-duality. The self-duality equations \rf{selfdualeqsgene}, for the static theory \rf{staticenergyfirstquasisd}, differs from the self-duality equations \rf{selfdualeqs} for the static theory \rf{staticenergy1} by the replacement $m_0\,e_0\rightarrow m_0\,e_0/\alpha$, with $\alpha$ given by \rf{alpha}. Note from \rf{alpha} that $\alpha=1$ implies $\beta_{{\cal V}}=0$, and so the absence of a potential term. On the other hand, the limit $\alpha\rightarrow 0$, corresponds to  strong coupling, i.e. $\beta_{{\cal V}}\rightarrow \infty$. In addition, the lower bound on the static energy, saturated by the self-dual solutions, grows monotonically with the increase of the potential strength.   Note that the $U$-fields are still totally free, as the $h$-fields still act as spectators. Indeed, given a $U$-field configuration, and so a $\tau$ matrix, the $h$-fields get determined in terms of $U$ by the equation \rf{ansatzh}.  

 \section{The second type of quasi-self-dual model}
\label{sec:second}
\setcounter{equation}{0}

We now consider the static theory $E=E_1+E_2$,  with $E_1$ given by \rf{staticenergy1}, and $E_2$ by \rf{e2def}, assuming only the quasi-self-duality condition \rf{halfbpscond}. We shall we a holomorphic ansatz for the $U$-fields, involving a radial profile function $f\(r\)$, and a complex field $u$ depending upon the angles of the spherical polar coordinates.

\subsection{The holomorphic ansatz}
\label{sec:secondholomorphic}

In order to construct an ansatz for the full theory \rf{full} we shall use the decomposition of the $SU(2)$ group element $U$ in terms of a real scalar field $f$ and a complex scalar field $u$, together with its complex conjugate $\ub$, as follows \cite{mantonbook,skyrmejoaq,laf2017}
\br
U=W^{\dagger}\,e^{i\,f\,T_3}\,W
\qquad\qquad 
{\rm with} \qquad\qquad  
W=\frac{1}{\sqrt{1+\u2}}\(
\begin{array}{cc}
1& i\,u\\
i\, \ub&1
\end{array}\)
\lab{udecompostion}
\er
Through \rf{udecompostion} the Maurer-Cartan can be writen as 
\be
R_i=R_i^a\,T_a=i\,\partial_{\mu}U\,U^{\dagger}=-G\, \Sigma_{i}\,G^{\dagger}\qquad \qquad{\rm with}\qquad\qquad  
G= W^{\dagger}\,e^{i\,f\,T_3/2}
\lab{usigmarel}
\ee
with 
\br
\Sigma_{i}
&=&\partial_{i}f\,T_3+\frac{2\,\sin\(f/2\)}{1+\u2}\left[i\,\partial_{i} \(u-{\bar u}\)\,T_{1}- \partial_{i}\(u+ {\bar u}\)\,T_{2}\right]
\lab{sigmamudef}
\er
From  \rf{usigmarel} we have that the matrix $\tau$, defined in \rf{taudef}, becomes
\br
\tau_{ab}= \trace\(\Sigma_i\,G^{\dagger}\,T_a\,G\)\,\trace\(\Sigma_i\,G^{\dagger}\,T_b\,G\)=
d^T_{ac}\(G^{\dagger}\)\, \trace\(\Sigma_i\,T_c\)\,\trace\(\Sigma_i\,T_d\)\, d_{db}\(G^{\dagger}\)
\lab{pretauholo}
\er
where $d^T\(G^{\dagger}\)=d\(G\)=d\(W^{\dagger}\)\,d\(e^{i\,f\,T_3/2}\)$ is the adjoint representation of $G^{\dagger}$, which using \rf{adjoint} and \rf{usigmarel} gives
\br
d\(e^{i\,f\,T_3/2}\)=\left(
\begin{array}{ccc}
 \cos \frac{f}{2} & \sin \frac{f}{2} & 0 \\
 -\sin \frac{f}{2} & \cos \frac{f}{2} & 0 \\
 0 & 0 & 1 \\
\end{array}
\right)
\er
and
\br
d\(W^{\dagger}\)=\frac{1}{1+\u2}\,\left(
\begin{array}{ccc}
 \frac{1}{2} \left(2+u^2+\ub^2\right) & \frac{1}{2} i \left(u^2-\ub^2\right) & i (u-\ub) \\
 \frac{1}{2} i \left(u^2-\ub^2\right) & \frac{1}{2} \left(2-u^2-\ub^2\right) & -(u+\ub ) \\
 -i (u-\ub) & u+\ub & 1-\u2  \\
\end{array}
\right)
\er

We now use spherical coordinates, but instead of using the polar and azimuthal angles we stereographic project the two sphere on a plane and parameterize that plane by a complex coordinate $w$, together with its complex conjugate $\wb$. So, we have the coordinate transformation
\be
x_1=r\;\frac{-i\(w-\wb\)}{1+\w2}\;;\qquad \qquad x_2=r\;\frac{\(w+\wb\)}{1+\w2}\;;\qquad \qquad
x_3=r\;\frac{\w2-1}{1+\w2}
\lab{wcartesiandef}
\ee
where $r$ is the radial distance. The  Euclidean space metric becomes 
\be
ds^2=dr^2+\frac{4\,r^2}{\(1+\w2\)^2}\,dw\,d\wb
\lab{holometric}
\ee
and so 
\be
d^3x=\sqrt{-g}\,\,dr\,dw\,d{\bar w}\;;\qquad\qquad\qquad   \sqrt{-g}=\frac{2\,r^2}{\(1+\w2\)^2}
\lab{volumeelement}
\ee
We now use the holomorphic ansatz for the $SU(2)$-fields defined by
\be
f\equiv f\(r\)\;;\qquad\qquad \qquad u\equiv u\(w\)\;;\qquad\qquad \qquad \ub\equiv \ub\(\wb\)
\lab{holoansatz}
\ee
where $u\(w\)$ is a map between two-spheres $(S^2)$. However, for the $u\(w\)$-field to be a well defined map between two-spheres it has to be a ratio of two polynomials $p_1$ and $p_2$, with no commum roots, i.e. the so-called rational map \cite{mantonbook,rational1, rational2}
\be
u\(w\)=\frac{p_1\(w\)}{p_2\(w\)}
\lab{rationalmap}
\ee
A well-known feature of the rational map \rf{rationalmap} is that its algebraic degree defined as the highest power of $w$ in either of the polynomials $p_1$ and $p_2$, corresponds exactly to its topological degree $n$, which can be writen in the integral representation as 
\be
n=\frac{1}{4\,\pi}\int d\,\Omega\,q =\frac{i}{2\,\pi}\int dw\wedge d\wb\,\frac{\mid\,p_2\,\partial_w p_1-p_1\,\partial_w p_2\mid^2}{\(\mid p_1\mid^2+\mid p_2\mid^2\)^2} \lab{n}
\ee
where $\Omega$ is the solid angle, and we use $d \Omega = \frac{2\,i \,dw\,\wedge \,d\wb }{\(1+ \mid w \mid^2\)^2} $ and  the follow definition
\be
q\equiv \frac{\(1+\w2\)^2}{\(1+\u2\)^2}\, \partial_w u\, \partial_{{\bar w}} {\bar u}
\lab{qdef}
\ee 
The topological charge density $\rho$ of \rf{topcharge} can be written using \rf{udecompostion}, \rf{holoansatz} and \rf{qdef} as   
\be
\rho\equiv \frac{i}{48\,\pi^2}\,\ve_{ijk}\,\trace\(R_i\,R_j\,R_k\)=-\frac{f'(r)}{4\,\pi^2}\,\frac{\sin^2\(f(r)/2\)}{r^2} \, q \lab{rho}
\ee
and so due to \rf{n} the topological charge \rf{topcharge} becomes 
\be
Q= \frac{\left[f-\sin f\right]_{f(\infty)}^{f(0)}}{2\,\pi}\,n \lab{fullq}
\ee
Note that due to \rf{rho} we get that ${\rm sign}\(Q\,f'\)=-1$ and so \rf{oposites} leads to
\be {\rm sign}\(Q\)=-{\rm sign}\(\lambda\)=-{\rm sign}\(f'\) \lab{signals}\ee

As a consequence of the holomorphic ansatz \rf{holoansatz} the matrix $\trace\(\Sigma_i\,T_a\)\,\trace\(\Sigma_i\,T_b\)$ becomes diagonal. 
Indeed, from \rf{sigmamudef}, \rf{holometric} and \rf{holoansatz}, one gets that
\be
\(\tau_D\)_{ab}\equiv \trace\(\Sigma_i\,T_a\)\,\trace\(\Sigma_i\,T_b\) = \omega_a\, \delta_{ab} \lab{tau}
\ee
with
\be
\omega_1=\omega_2= \frac{4\, \sin^2\(f/2\)}{r^2} \, q \qquad\qquad \qquad\qquad\qquad \omega_3= \(f^{\prime}\)^2 
\lab{omegaholo}
\ee 
where prime denotes derivatives with respect to $r$. Comparing \rf{diagonalhtau} and \rf{pretauholo} we then conclude that
\be
N=d^T\(G^{\dagger}\)=d\(G\)= d\(W^{\dagger}\)\,d\(e^{i\,f\,T_3/2}\)
\lab{nexpression}
\ee

\subsection{The Euler-Lagrange equations}
\label{sec:elequationsholomorphic}

We start the analysis of the Euler-Lagrange equations, in the holomorphic ansatz, by noticing that, if one considers $u$, ${\bar u}$, $\partial_{ w} { u}$, and $\partial_{\bar w} {\bar u}$, as independent variables, then the quantity $q$ defined in   \rf{qdef}, satisfies  
\be
\frac{\delta\,q}{\delta\,u}-\(1+\mid w\mid^2\)^2\,\partial_w\(\frac{1}{\(1+\mid w\mid^2\)^2}\frac{\delta\,q}{\delta\partial_w u}\)=0
\lab{niceqeleq}
\ee
together with its complex conjugate. 

From \rf{omegaholo} we have that $\omega_3$ depends only on the radial profile function $f\(r\)$, and $\omega_1$ and $\omega_2$ depend upon $u$ and ${\bar u}$ through $q$ only, and they are linear in $q$. Therefore, from \rf{deltaue1} we observe that the variation of $E_1$ with respect the $u$-field is
\be
\delta^{(u)} E_1=\frac{1}{2\,e_0^2}\, \int d^3x\,{\hat \omega}\(r\)\left[ m_0^2\,e_0^2\, \(\vp_1+\vp_2\)+\omega_3\(r\)\(\frac{1}{\vp_1}+\frac{1}{\vp_2}\)+2\,{\hat \omega}\(r\)\,\frac{q}{\vp_3}\right]\,\delta^{(u)}q
\lab{preelue1}
\ee
where, following \rf{omegaholo}, we have defined 
\be
{\hat \omega}\(r\)\equiv  \frac{4\, \sin^2\(f/2\)}{r^2}
\lab{omegahatdef}
\ee
Consequently, if we consider the ansatz
\be
\vp_1=\vp_1\(r\);\qquad \qquad\vp_2=\vp_2\(r\);\qquad \qquad\vp_3={\hat \vp}_3\(r\)\, q
\lab{preansatzvp}
\ee 
we get, using \rf{volumeelement}, that \rf{preelue1} vanishes as a consequence of \rf{niceqeleq}. For the same reasons one gets that $\delta^{({\bar u})} E_1=0$. 
From \rf{udecompostion} we see that ${\rm Tr}U$ does not depend upon the fields $u$ and ${\bar u}$. Therefore, from \rf{deltaue2} we get that
\be
\delta^{(u)} E_2=\delta^{({\bar u})} E_2=0
\ee
So, the conditions \rf{preansatzvp} are sufficient for the Euler-Lagrange equations, associated to the $u$ and ${\bar u}$ fields, to be satisfied, within the holomorphic ansatz \rf{holoansatz}. 

Using \rf{omegahatdef} and \rf{preansatzvp} one gets from \rf{deltaue1}  that
\br
\delta^{(f)}E_1&=&\frac{1}{2\,e_0^2}\, \int d^3x\,\left[m_0^2\,e_0^2\, \(\vp_1+\vp_2\)\,\delta^{(f)}{\hat\omega}+\(\frac{1}{\vp_1}+\frac{1}{\vp_2}\)\,\delta^{(f)}\({\hat\omega}\,\omega_3\)
\right.\nonumber\\
&+&\left. m_0^2\,e_0^2\;{\hat \vp}_3\,\delta^{(f)}\omega_3+ \frac{\delta^{(f)}{\hat\omega}^2}{{\hat\vp}_3}\right]\,q
\er
As $q$ factors out, one observes that the variation $E_1$ with respect to the profile function $f$ leads to a radial equation for it. However, from \rf{deltaue2} and \rf{udecompostion} one gets that 
\be
\delta^{(f)}E_2=\frac{\beta_3^2}{2}\,\int d^3x \;\sin\(\frac{f}{2}\)
\lab{deltaue2}
\ee
Therefore, for $\beta_3\neq 0$, one has to impose that $q$ must be a constant, in order to get a radial equation for $f$. 

Let us now analyze the Euler-Lagrange equations for the $M$-fields given in \rf{eqforM}. Our quasi-self-dual condition \rf{halfbpscond} requires $M=N$, and so \rf{eqforM} becomes in fact equations for the $U$-fields. We do not want the $U$-fields to be subjected to additional equations, besides their own Euler-Lagrange equations. Therefore, we want  \rf{eqforM} to be solved automatically by the holomorphic ansatz, supplemented by some extra conditions. From \rf{nexpression} and the holomorphic ansatz \rf{holoansatz} for the $U$-fields we obtain 
\br
N^T\,\partial_r N&=& \frac{i}{2} \, f^{\prime} \, d\(T_3\)
\nonumber\\
N^T\,\partial_{w} N &=&\frac{1}{1+\u2}\,\left[- {\bar u} \partial_w u\,d\(T_3\)-i\,e^{-i\,f/2}\,\partial_w u\,d\(T_{1}+i\,T_2\)\right]
\lab{NTderN}
\\
N^T\,\partial_{{\bar w}} N &=&\frac{1}{1+\u2}\,\left[ u\, \partial_{\bar w} {\bar u}\,d\(T_3\) -i\,e^{i\,f/2}\,\partial_{\bar w} {\bar u}\,d\(T_{1}-i\,T_2\)\right]
\nonumber
\er
where we have used the fact that $d_{ab}\(T_c\)=i\,\ve_{acb}$.  Therefore,  using \rf{doublecomhd}, we get that
 \br
\sbr{h_D}{\sbr{h_D}{N^T\,\partial_r N}}&=& \frac{i}{2} \, f^{\prime} \, \(\vp_1-\vp_2\)^2\,d\(T_3\)
\nonumber\\
\sbr{h_D}{\sbr{h_D}{N^T\,\partial_{w} N }}&=&\frac{1}{1+\u2}\,\left[- {\bar u} \partial_w u\,\(\vp_1-\vp_2\)^2\,d\(T_3\) \lab{comut}
\right. \\
&-&\left. i\,e^{-i\,f/2}\,\partial_w u\,\left[\(\vp_2-\vp_3\)^2\,d\(T_{1}\)+i\,\(\vp_1-\vp_3\)^2\,d\(T_2\)\right]\right]
\nonumber\\
\sbr{h_D}{\sbr{h_D}{N^T\,\partial_{{\bar w}} N}} &=&\frac{1}{1+\u2}\,\left[ u\, \partial_{\bar w} {\bar u}\,\(\vp_1-\vp_2\)^2\,d\(T_3\) 
\right. \nonumber\\
&-&\left.i\,e^{i\,f/2}\,\partial_{\bar w} {\bar u}\,\left[\(\vp_2-\vp_3\)^2\,d\(T_{1}\)-i\,\(\vp_1-\vp_3\)^2\,d\(T_2\)\right]\right]
\nonumber
\er
It then follows that 
\br
\partial_r\sbr{h_D}{\sbr{h_D}{N^T\,\partial_r N}}+\sbr{N^T\,\partial_r N}{\sbr{h_D}{\sbr{h_D}{N^T\,\partial_r N}}}=
\frac{i}{2} \,\( f^{\prime} \, \(\vp_1-\vp_2\)^2\)^{\prime}\,d\(T_3\)
\nonumber
\er
But that involves first and second derivatives of the profile function $f$, which can not be canceled by the remaining terms of \rf{eqforM}. Therefore, we shall impose, besides \rf{preansatzvp},  the condition $\vp_1\(r\)=\vp_2\(r\)$. One can then check that all the terms in \rf{eqforM} vanish except for those involving $w$ and ${\bar w}$ derivatives of the $\vp_3$ field. 
Considering the form of $\vp_3$, given in \rf{preansatzvp}, those derivatives do not cancel each other unless we assume that $q$ is constant. However, as shown in Appendix \ref{app:proof} the only rational maps \rf{rationalmap} that leads to a constant value of $q$ have the form $u=e^{i\,\alpha}\,w$ or $u=\frac{\beta\(w-\mid \beta \mid^{-1}\,e^{i\alpha}\)}{w+\mid \beta \mid\,e^{i\alpha}}$, where $\alpha$ is a real constant contained in the interval $\left[0,\,2\pi\)$ and $\beta$ is a arbitrary complex constant with $\beta \neq 0$. Note that  both of these rational maps leads to $q=1$. Therefore, we are lead to consider the following ansatz for the $\vp$-fields 
\be
\vp_1=\vp_2\equiv \vp_1\(r\);\qquad\qquad\qquad \vp_3= \vp_3\(r\)
\lab{varphiansatz}
\ee
and for the $u$-fields 
\be
u=e^{i\,\alpha}\,w\;\qquad\quad {\rm or}\qquad\quad u=\frac{\beta\(w-\mid \beta \mid^{-1}\,e^{i\alpha}\)}{w+\mid \beta \mid\,e^{i\alpha}};\qquad\qquad \mbox{\rm and so}\qquad\qquad q=1
\lab{q=1ansatz}
\ee
Note that imposing that $q$ must be constant is equivalent to imposing that the topological charge density inside the holomorphic ansatz, as given in \rf{rho}, must have radial symmetry. For the rational maps \rf{q=1ansatz}, which have topological degree $n=1$, the topological charge \rf{fullq} becomes
\be
Q= \frac{\left[f-\sin f\right]_{f(\infty)}^{f(0)}}{2\,\pi} \lab{smallq}
\lab{q=1topcharge}
\ee

Summarizing, using the holomorphic ansatz \rf{holoansatz} together with the conditions \rf{varphiansatz} and \rf{q=1ansatz} we get that the Euler-Lagrange equations for the $M$,  $u$ and ${\bar u}$ fields are automatically satisfied. We are then left with three radial equations which are the Euler-Lagrange equations for the profile function $f$ and for the $\vp$-fields. 

The Euler-Lagrange equation for $f$ is given by 
\br
&&m_0^2\left[\frac{1}{r^2}\partial_r\(r^2\,\vp_3\,f^{\prime}\)-2\,\vp_1\,\frac{\sin f}{r^2}\right]-\frac{\beta_3^2}{2}\,\sin\(f/2\)
\lab{eqforffinal}\\
&&+\frac{1}{e_0^2}\left[\frac{1}{r^2}\partial_r\(\frac{8\,\sin^2\(f/2\)}{\vp_1}\,f^{\prime}\)-\frac{16\,\sin^3\(f/2\)\,\cos\(f/2\)}{r^4\,\vp_3}-\frac{2\,\sin f\;\(f^{\prime}\)^2}{r^2\,\vp_1}
\right]=0
\nonumber
\er
The Euler Lagrange equations for the $\vp_1=\vp_2$ and $\vp_3$ fields are respectively
\br
\frac{\mu_0^2}{r^2}\left[\kappa_1\partial_r\(r^2\vp_1^{\prime}\)-\kappa_4\,2\(\vp_1-\vp_3\)\right]-\frac{\delta\,{\cal V}}{\delta\,\vp_1}
-m_0^2\,\frac{2\,\sin^2\(f/2\)}{r^2}\left[1-\frac{1}{m_0^2\,e_0^2}\,\frac{\(f^{\prime}\)^2}{\vp_1^2}\right]=0
\lab{eqforphifinal1}
\er
and
\br
\frac{\mu_0^2}{r^2}\left[\kappa_3\partial_r\(r^2\vp_3^{\prime}\)+\kappa_4\,4\(\vp_1-\vp_3\)\right]-\frac{\delta\,{\cal V}}{\delta\,\vp_3}
-\frac{m_0^2}{2}\left[\(f^{\prime}\)^2-\frac{16\,\sin^4\(f/2\)}{m_0^2\,e_0^2\,r^4\,\vp_3^2}\right]=0
\lab{eqforphifinal3}
\er
Due to the condition \rf{varphiansatz} we had to assume that the potential ${\cal V}$ is symmetric under the exchange $\vp_1\leftrightarrow \vp_2$, and that the coupling constants $\kappa_1$ and $\kappa_2$, introduced in \rf{e2def}, are the same. 

In the next section we show how to solve numerically those three radial equations.

\subsection{Numerical solutions for a quadratic potential ${\cal V}$}
\label{sec:numerics}

Consider the static sector of the theory \rf{full} with ${\cal V}=\frac{\beta_1^2}{2}\,{\rm Tr}\,h^2=\frac{\beta_1^2}{2}\,\sum_{a=1}^3\vp_a^2$,  and $\kappa_{\alpha}=1$, $\alpha=1,2,3,4$. As we are working with the ansatz\"e \rf{halfbpscond}, \rf{holoansatz}, \rf{varphiansatz} and \rf{q=1ansatz}, we shall be concerned with configurations of unity topological charge only. Therefore, we shall  measure the energy in units of  $48\,\pi^2\,\frac{\mid m_0 \mid}{\mid e_0\mid}$. That means that the BPS energy \rf{staticenergy1bps} of the self-dual configurations \rf{hbps} becomes $E_1^{\rm BPS}=1$, for $Q=1$.  We shall measure length in units of $\mu_0^2/(m_0^3 \,e_0)$, and rescale the  $h$-fields, and so the $\vp$-fields,  by the dimensionless factor  $m_0^2/\mu_0^2$. Therefore, using \rf{staticenergy1} and \rf{e2def} the total static energy can be rewriten, in terms of the new units, as
\be
E= \frac{1}{96\,\pi^2}\int d^3 x\left[h_{ab}\,R^a_i\,R^b_i+\frac{1}{2}\, h^{-1}_{ab}\,H^a_{ij}\,H^{b}_{ij}
+{\rm Tr}\(\partial_i h\)^2+\sigma_1\, {\rm Tr}\,\(h^2\)
+\sigma_2\, {\rm Tr}\(\one - U\)\right] 
\lab{modelenergy2}
\ee
with 
\be
\sigma_1=\frac{\mu_0^2\,\beta_1^2}{m_0^6e_0^2}\;;\qquad\qquad\qquad \sigma_2=\frac{\mu_0^2\,\beta_3^2}{m_0^6e_0^2} \lab{sigs}
\ee
The Euler-Lagrange equations \rf{eqforffinal}, \rf{eqforphifinal1} and \rf{eqforphifinal3} becomes respectively
\br
\Delta E_f&\equiv& \frac{1}{r^2}\left[\partial_{r}\(r^2\, A\)+2\, B\, \sin\(f\)\right]+\frac{\sigma_2}{2}\, \sin\(f/2\)=0, \lab{deltaef}\\
\Delta E_{\vp_1}&\equiv&\vp_1^{\prime\prime}+\frac{2}{r}\,\vp_1^{\prime}-\frac{2}{r^2}\(\vp_1-\vp_3\) - 
\sigma_1\,\vp_1 -\frac{2}{r^2}\sin^2\(f/2\)\left[1-\frac{\(\pa_r f\)^2}{\vp_1^2}\right]=0,
\lab{deltaep1} \\
\Delta E_{\vp_3}&\equiv& \vp_3^{\prime\prime}+\frac{2}{r}\,\vp_3^{\prime}+\frac{4}{r^2}\(\vp_1-\vp_3\) - 
\sigma_1\,\vp_3 -\frac{1}{2}\,\(f^{\prime}\)^2\,\left[1-\frac{16\,\sin^4\(f/2\)}{r^4\,\vp_3^2\,\(\pa_r f\)^2}\right]=0 ,\quad\qquad
\lab{deltaep3} 
\er
where 
\be
A\equiv-\partial_{r}f\left[\vp_3+\frac{8\,\sin^2\(f/2\)}{r^2\,\vp_1}\right],\qquad 
B\equiv\vp_1\left[1+\frac{\(\partial_{r}f\)^2}{\vp_1^2}+\frac{4\,\sin^2\(f/2\)}{r^2\,\vp_1\,\vp_3}\right] .
\ee

Inside ansatzë \rf{halfbpscond}, \rf{holoansatz}, \rf{varphiansatz} and \rf{q=1ansatz} the static energy \rf{modelenergy2} is reduced to
\br E=E_1+E_2;\qquad\qquad E_1 =\mathcal{E}_2+\mathcal{E}_4;\qquad\qquad E_2 =\mathcal{E}_h+\mathcal{E}_{\sigma_1}+\mathcal{E}_{\sigma_2} \lab{energiessum}\er
with 
\br
\mathcal{E}_{2}&\equiv& \frac{1}{96\,\pi^2}\,\int d^3x\, h_{ab}\,R^a_i\,R^b_i
=\frac{1}{12\,\pi}\int dr \,r^2\(\frac{{\vp}_3 \,\(\pa_r f\)^2}{2}+\vp_1\frac{4\,\sin^2\frac{f}{2}}{r^2}\) \nonumber\\
\mathcal{E}_{4}&\equiv& \frac{1}{192\,\pi^2}\,\int d^3x\,  h^{-1}_{ab}\,H^a_{ij}\,H^{b}_{ij}=\frac{1}{12\pi}\int dr \,r^2\left[\frac{4\,\sin^2\frac{f}{2}}{r^2}\(\frac{\(\pa_r f\)^2}{\vp_1}+\frac{2\,\sin^2\frac{f}{2}}{r^2 \,{\vp}_3}\) \right] \nonumber\\
\mathcal{E}_{h}&\equiv& \frac{1}{96\,\pi^2}\,\int d^3x\, {\rm Tr}\, \(\pa_i h\)^2
=\frac{1}{12\,\pi}\,\int dr \,r^2\left[\(\pa_r\vp_1\)^2+\frac{\(\pa_r\vp_3\)^2}{2}+\frac{2\,\(\vp_1 -{\vp}_3 \)^2}{r^2}\right]\lab{staticenergies}\\
\mathcal{E}_{\sigma_1}&\equiv& \frac{\sigma_1}{96\,\pi^2}\,\int d^3x\, {\rm Tr}\,\( h^2\)=\frac{\sigma_1}{12\,\pi}\int dr \,r^2\(\vp_1^2+\frac{{\vp}_3^2}{2}\) \nonumber\\
\mathcal{E}_{\sigma_2}&\equiv& \frac{\sigma_2}{96\,\pi^2}\,\int d^3x \, {\rm Tr}\(\one -U\)=\frac{\sigma_2}{12\,\pi}\int dr\,r^2\(1-\cos\frac{f}{2}\)  \nonumber
\er

The stability of the solutions of \rf{deltaef}-\rf{deltaep3}  under the scale Derrick's argument \cite{derrick,coleman2} imposes relations only between the terms of $E_2$,  since $E_1$ is conformal invariant in three spatial dimensions. Indeed, since the $h$-fields have conformal weight $-1$ by the scaling transformation $x\rightarrow \alpha\,x$ these fields must transform as $h\rightarrow \alpha^{-1}\,h$, and so the $E_2$ terms of \rf{energiessum} transforms as $\mathcal{E}_{h}\rightarrow \alpha^{-1}\,\mathcal{E}_{h}$, $\mathcal{E}_{\sigma_1}\rightarrow \alpha\,\mathcal{E}_{\sigma_1}$ and $\mathcal{E}_{\sigma_2}\rightarrow \alpha^3\,\mathcal{E}_{\sigma_2}$. Therefore, the stable solutions under the Derrick's argument need to satisfy
\br -\mathcal{E}_h+\mathcal{E}_{\sigma_1}+3\,\mathcal{E}_{\sigma_2}&=&0;\qquad\qquad\qquad \mathcal{E}_h+3\,\mathcal{E}_{\sigma_2}>0\lab{derrick1}\er
The inequality of \rf{derrick1} is automatically satisfied and the first relation imposes that the dimensionless quantity 
\br {\rm Derrick} \equiv \frac{\left|-\mathcal{E}_h+\mathcal{E}_{\sigma_1}+3\,\mathcal{E}_{\sigma_2}\right|}{E_2} \lab{derrick3}\er
must be zero. Note that the term $E_2$ in the denominator of the l.h.s. of \rf{derrick3} prevents unstable solutions of \rf{deltaef}-\rf{deltaep3} from leading to small values of the quantity \rf{derrick3} under weak coupling regime $E_2\ll E_1$, where all the terms $\mathcal{E}_h$, $\mathcal{E}_{\sigma_1}$ and $\mathcal{E}_{\sigma_2}$ are small.

The simplest topological solutions that we can construct are those with Skyrme charge $Q=1$, and due to \rf{smallq} we shall impose the boundary conditons $f(0)=2\,\pi$ and $f(\infty)=0$. So, expanding \rf{deltaef}, \rf{deltaep1} and \rf{deltaep3} in Taylor series at $r =0$ we obtain
\br
\pa_r^2 f(0)&=&\pa^4_r f(0)=0; \qquad\quad \pa_r\vp_a(0)=\pa_r^3\vp_a(0)=0;\qquad\quad \vp_1(0)=\vp_3(0) \lab{expansion0}\er 
with $a=1,\,3$. 
We use the gradient flow method with adaptive step size to minimize the static energy \rf{energiessum} and to get the solutions of \rf{deltaef}-\rf{deltaep3} with $Q=1$, as described in the Appendix \ref{app:numerics}. The coordinate $r$ lies  in the interval $[0\,,r_{\rm max}]$, where $r_{\rm max}$ is the size of the lattice. Table \ref{table1} shows the energies \rf{staticenergies} corresponding to the solutions of \rf{deltaef}-\rf{deltaep3} for some pairs of values of $\sigma_1$ and $\sigma_2$ labeled by an index $N_c$. The highest value of \rf{derrick3} is $3.11\times 10 ^{-3}$, which means that on all numerical solution presented in the Tables \ref{table1} and \ref{table2} the term $\left|-\mathcal{E}_h+\mathcal{E}_{\sigma_1}+3\,\mathcal{E}_{\sigma_2}\right|$ of \rf{derrick1} is equal to or less than $0.311\,\%$ of $E_2$. So, the relation \rf{derrick1} imposed by the Derrick's scale argument are satisfied with a quite good precision. 

\begin{figure}[H]
\centering
\includegraphics[scale=0.63,angle=0]{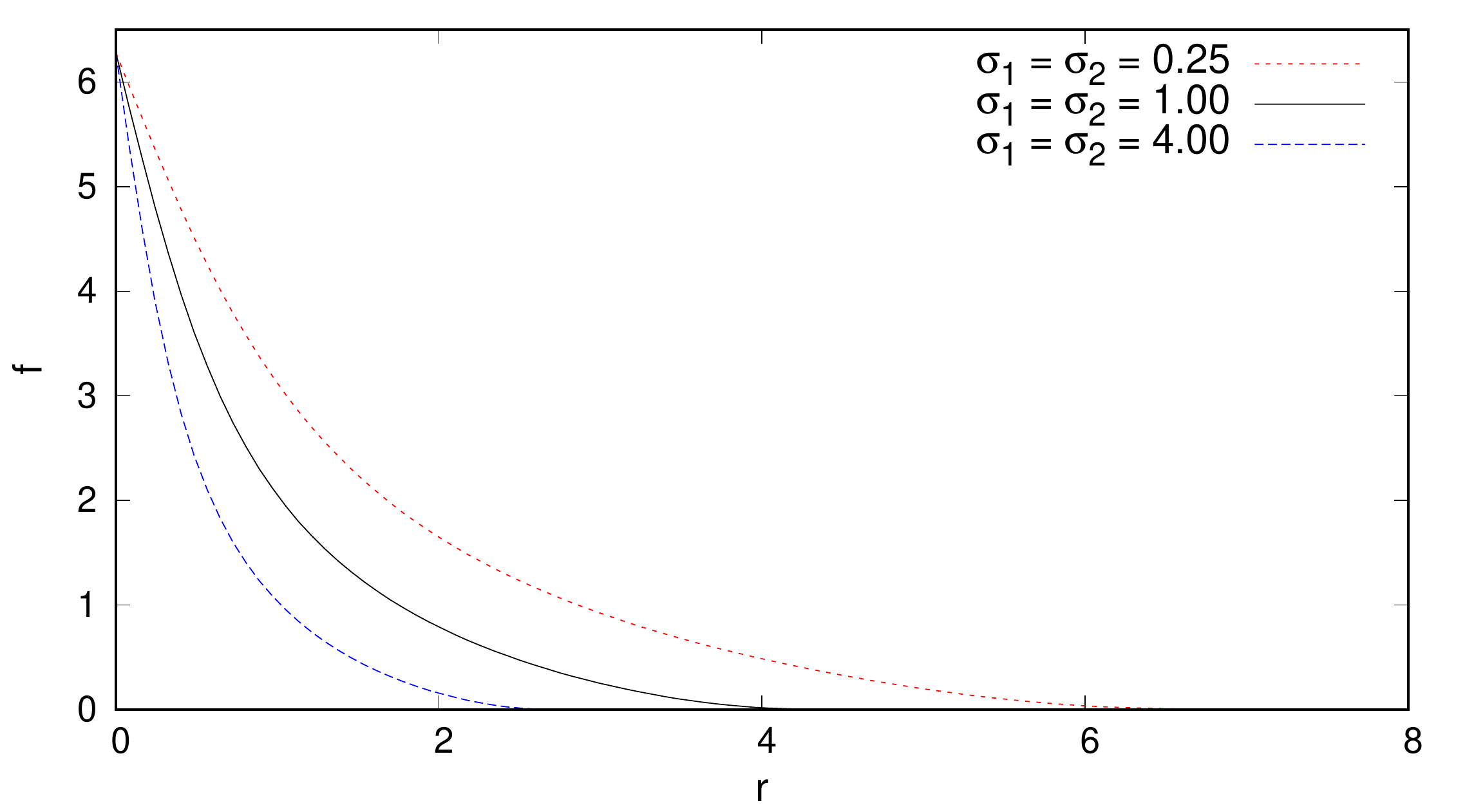}
\caption{The $f(r)$-field solution of \rf{deltaef}-\rf{deltaep3} corresponding to $Q=1$ for $\sigma_1=\sigma_2 = 0.25,\,1.00,\,4.00$.}
\label{fig1}
\end{figure}

\begin{figure}[H]
\centering
\includegraphics[scale=0.63,angle=0]{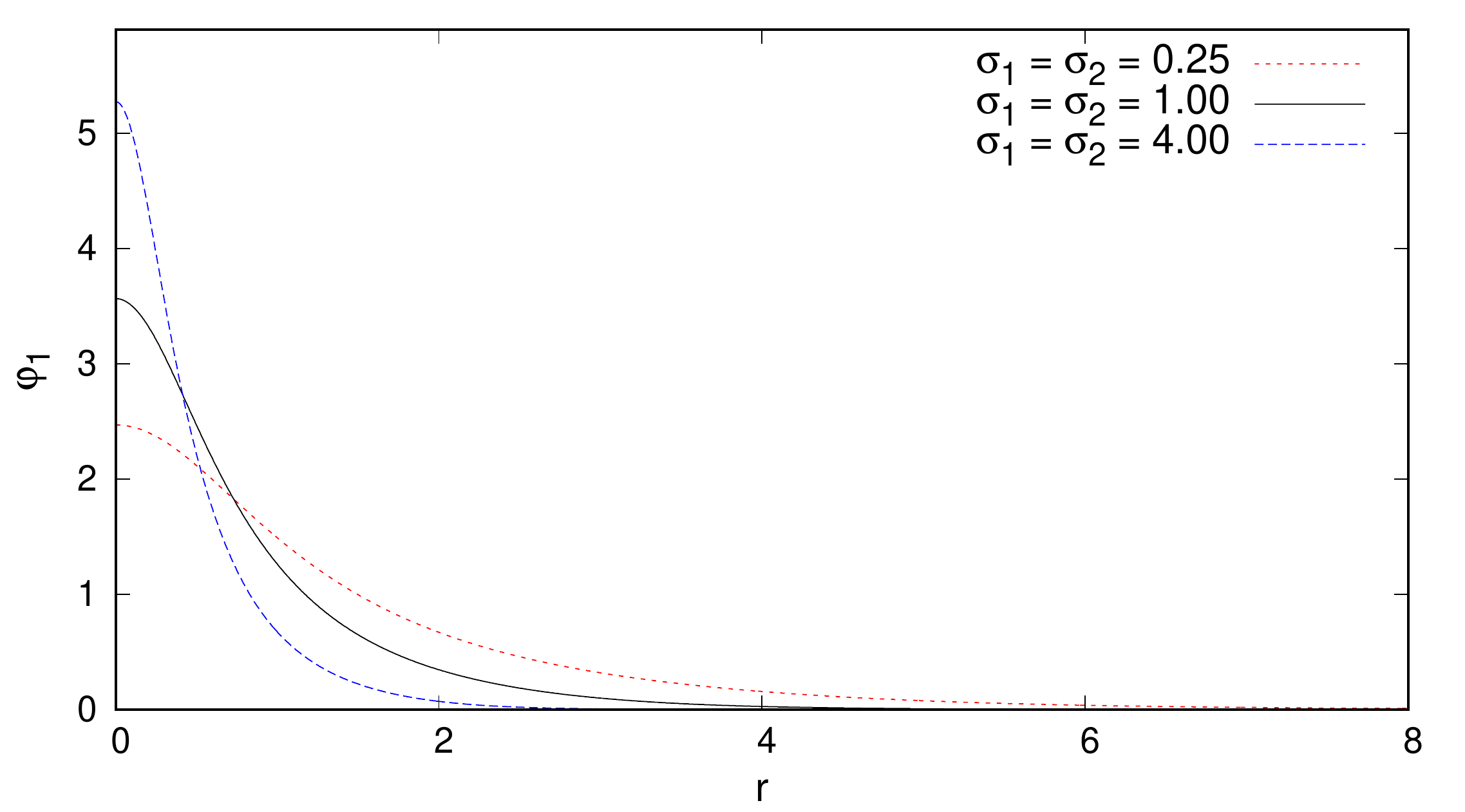}
\caption{The $\vp_1(r)$-field solution of \rf{deltaef}-\rf{deltaep3} corresponding to $Q=1$ for $\sigma_1=\sigma_2 = 0.25,\,1.00,\,4.00$.}
\label{fig2}
\end{figure}

\begin{figure}[H]
\centering
\includegraphics[scale=0.63,angle=0]{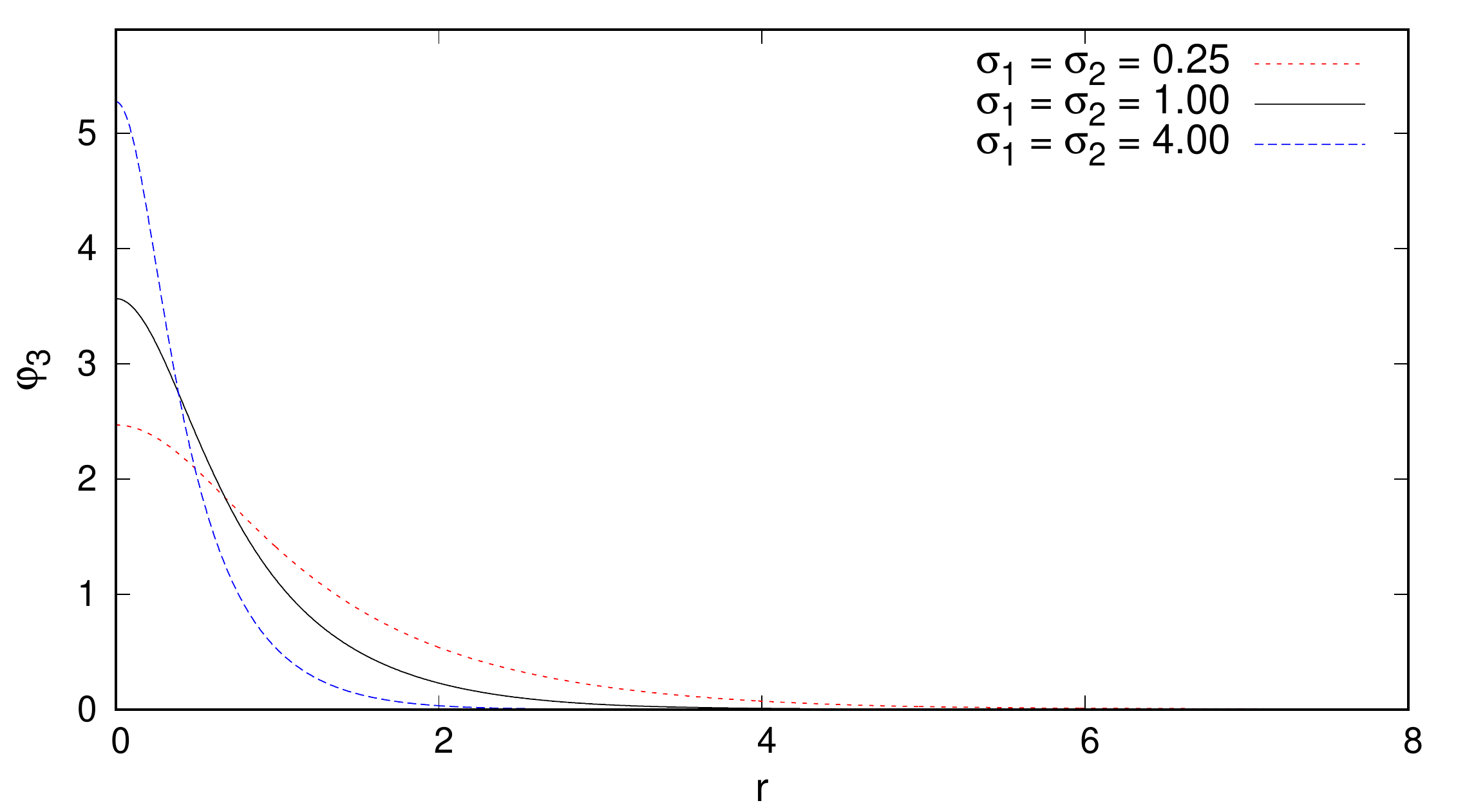}
\caption{The $\vp_3(r)$-field solution of \rf{deltaef}-\rf{deltaep3}, corresponding to $Q=1$ for $\sigma_1=\sigma_2 = 0.25,\,1.00,\,4.00$.}
\label{fig3}
\end{figure}

The numerical solutions of $f(r)$ and $\vp_a(r)$ obtained for all parameters of Table \ref{table1} are monotonically decreasing (see examples in the Figures \ref{fig1}-\ref{fig3}). The amplitude $\vp_a(0)$ of the eigenvalues of $h$ and the thickness $t$ for each of the fields, defined as the value of $r$ for which the field reaches half of its value at $r=0$, are given in the Table \ref{table2}. On the Table \ref{table2} and in the Figure \ref{fig1} we can see that the thickness of $f$ decreases when $\sigma_2$ grows and the $U$-fields becomes more massive. The same follows for the $\vp_a$-fields when $\sigma_1$ grows and the $h$-fields becomes more massive, but in contrast their amplitude increases (see Figures \ref{fig1} and \ref{fig3}). 

The quadratic and quartic term of \rf{staticenergy1} in the spatial derivatives becomes the same for the self-dual configurations \rf{hbps} (see \rf{relations2}), and so in the units defined above we must have $\mathcal{E}_2^{{\rm BPS}}=\mathcal{E}_4^{{\rm BPS}}= 1/2$. In additional, any solutions of \rf{deltaef}-\rf{deltaep3} must satisfies the Bogomolny bound $E_1 \geq E_1^{{\rm BPS}}=1$, which has its lower bound saturated by \rf{hbps} (see Table \ref{table2}). 
The self-dual solutions \rf{hbps} are conformally invariant in $\IR^3$ and possesses a infinite number of exact solutions for each value of $Q$, and so we can not directly compare their the shape of the BPS configurations with the solutions of \rf{deltaef}-\rf{deltaep3}. However, we can use the quantity $\(E_1-1\)$ and the ratio $\mathcal{E}_4/\mathcal{E}_2$ to mesure how far the static solutions of the full theory \rf{modelenergy2} are from the self-dual sector of the BPS Skyrme model \rf{modelintro}. Indeed, from Table \ref{table1} we see that these two quantities tend to increases, getting farther and farther from 1, when either $\sigma_1$ or $\sigma_2$ grow and are more sensitive to $\sigma_1$ than $\sigma_2$.

\begin{table}[H]
	\caption{The quantities \rf{energiessum}, \rf{staticenergies} and \rf{derrick3} associated with the solutions of the equations \rf{deltaef}-\rf{deltaep3} with $Q=1$ for some values of $\sigma_1$ and $\sigma_2$.}
		\label{table1}
    \centering
\begin{tabular}{cccccccccc}\hline
$N_c$ & $\sigma_1$ & $\sigma_2$ &	$E$ & Derrick & $\mathcal{E}_2$ & $\mathcal{E}_4$ & $\mathcal{E}_h$ & $\mathcal{E}_{\sigma_1}$ & $\mathcal{E}_{\sigma_2}$\\ \hline
$i$ & $ 0.25 $ & $ 0.25 $ & $ 1.26924 $ & $ 2.48 \times 10^{-3}$ & $ 0.35860 $ & $ 0.70973 $ & $ 0.12606 $ & $ 0.04950 $ & $ 0.02535 $ \\
$ii$ & $ 0.50 $ & $ 0.50 $ & $ 1.32700 $ & $ 3.11\times 10^{-3}  $ & $ 0.33712 $ & $ 0.75577 $ & $ 0.14222 $ & $ 0.06711 $ & $ 0.02479 $ \\ 
$iii$ & $ 1.00 $ & $ 1.00 $ & $ 1.39771 $ & $ 8\times 10^{-5} $ & $ 0.31419 $ & $ 0.81127 $ & $ 0.15982 $ & $ 0.08873 $ & $ 0.02370 $ \\
$iv$ & $ 2.00 $ & $ 2.00 $ & $ 1.48360 $ & $ 3.3\times 10^{-4} $ & $ 0.29045 $ & $ 0.87773 $ & $ 0.17945 $ & $ 0.11419 $ & $ 0.02179 $ \\
$v$ & $ 4.00 $ & $ 4.00 $ & $ 1.58693 $ & $ 1.3\times 10^{-4} $ & $ 0.26657 $ & $ 0.95620 $ & $ 0.20139 $ & $ 0.14342 $ & $ 0.01934 $  \\ 
$vi$ & $ 0.25 $ & $ 4.00 $ & $ 1.37545 $ & $ 1.1\times 10^{-4}$ & $ 0.34137 $ & $ 0.76872 $ & $ 0.18433 $ & $ 0.02934 $ & $ 0.05167 $ \\
$vii$ & $ 4.00 $ & $ 0.25 $ & $ 1.56171 $ & $ 7 \times 10^{-5} $ & $ 0.26615 $ & $ 0.95056 $ & $ 0.17531 $ & $ 0.16689 $ & $ 0.00280 $ \\ \hline
\end{tabular}
\end{table}

\begin{table}[H]
	\caption{The thicknesses of solutions with $Q=1$ of the equations \rf{deltaef}-\rf{deltaep3}, the size of the lattice $r_{\rm max.}$ and the quantities $\(E_1-1\) $ and $\mathcal{E}_4/\mathcal{E}_2$ for the values of $\sigma_1$ and $\sigma_2$ from Table \ref{table1}.}
		\label{table2}
    \centering
\begin{tabular}{cccccccc}\hline
$N_c$ & $i$ & $ii$ & $iii$ & $iv$ & $v$ & $vi$ & $vii$ \\\hline 
$t_f$ & $ 0.99040 $  & $ 0.77787 $  & $ 0.60444 $  & $ 0.46323 $  & $ 0.35029 $  & $ 0.53182 $  & $ 0.41235 $ \\
$t_{\vp_1}$ & $ 1.23552 $  & $ 0.96684 $  & $ 0.74737 $  & $ 0.56949 $  & $ 0.42818 $  & $ 0.71016 $  & $ 0.48892 $ \\
$t_{\vp_3}$ & $ 1.13680 $  & $ 0.89573 $  & $ 0.69698 $  & $ 0.53423 $  & $ 0.40373 $  & $ 0.64507 $  & $ 0.46551 $ \\
$r_{\rm max}$ & $ 25 $ & $ 25 $ & $ 14 $ & $ 14 $ & $ 14 $ & $ 25$ & $ 14 $  \\
$E_1-1$ &  $ 0.06832 $ &  $ 0.09289 $ &  $ 0.12547 $ &  $ 0.16817 $ &  $ 0.22277 $ & $0.11010 $ &  $ 0.21671 $ \\
$\mathcal{E}_4/\mathcal{E}_2$ &  $ 1.9792 $ & $ 2.2418 $ & $ 2.5821 $ & $ 3.0220 $ & $ 3.5870 $ & $ 2.25188 $ & $ 3.5716 $ \\
\hline
\end{tabular}
\end{table}

\section{Conclusion}
\label{sec:conclusion}
\setcounter{equation}{0}

We have proposed extensions of the Skyrme model \rf{modelintro} that allows the breaking of its self-dual sector in a soft manner. The self-duality equations \rf{selfdualeqs} impose that the matrix $h$ must be proportional to the matrix $\tau$, as shown in \rf{hbps}. Therefore, the two matrices are diagonalized by the same orthogonal matrix $M$, see \rf{diagonalhtau}, and their eigenvalues are related by \rf{useful}. We extend  the theory \rf{modelintro} by introducing kinetic and potential terms for the $h$-fields, and impose that the matrices $h$ and $\tau$ should still be diagonalised by the same orthogonal matrix $M$. That is our conditions \rf{halfbpscond}, which we call quasi-self-duality equations. 

We study two distinct cases of the breaking of the self-duality equations. The first one comes from the observation that by imposing algebraic relations among the eigenvalues of the matrices $h$ and $\tau$, given in \rf{funnyrel}, one gets that the part of the Euler-Lagrange equations associated to the $U$-fields, coming from $E_1$, given in \rf{staticenergy1}, is automatically satisfied, since the variation of $E_1$ becomes proportional to the variation of the topological charge, as shown in \rf{vareiequalvartopcharge}. The other observation is that the variation of $E_1$ with respect to the $\vp$-fields is proportional solely to the variation of  ${\rm det}\, h$. Therefore, choosing $E_2$ to contain just a potential term proportional to ${\rm det}\, h$, one solves the Euler-Lagrange equations for the $\vp$ and $U$-fields. We are then led to the theory  \rf{staticenergyfirstquasisd}, which is shown to possess an exact self-dual sector. The corresponding self-duality equations are given in \rf{selfdualeqsgene}, and they differ from the original self-duality equations \rf{selfdualeqs} by a constant $\alpha$ which is a monotonically decreasing function of the strength of the potential proportional to  ${\rm det}\, h$. The theory \rf{staticenergyfirstquasisd} has the same  global symmetry $SU(2)_L\otimes SU(2)_R$, as \rf{modelintro}, and it is also conformally invariant in the three dimensional spatial submanifold $\IR^3$. In addition, the static energy of the self-dual solutions is proportional to the topological charge $Q$, and the proportionality constant grows with the strength of the potential ${\rm det}\, h$. As in the original theory \rf{modelintro}, the $h$-fields act as spectators in the sense that, given a configuration for the $U$-fields, they adjust themselves to solve the self-duality equations.  It is remarkable that a theory like \rf{staticenergyfirstquasisd} possesses an exact self-dual sector. That may lead to new interesting applications, specially for nuclear matter as done in \cite{usfalse} for the theory \rf{modelintro}.

The second way of breaking the self-duality of the theory \rf{modelintro}, but respecting the quasi-self-duality equations  \rf{halfbpscond}, is by introducing kinetic and potential terms for the $\vp$-fields. In order to study such a case we use the holomorphic ansatz for the $U$-fields, given in \rf{udecompostion} and \rf{holoansatz}. As a consequence of \rf{halfbpscond},  the Euler-Lagrange equations for the $M$-fields become extra conditions for the $U$-fields to satisfy. In order to avoid such strongly restricting conditions, we solve the equations for the $M$-fields  by imposing  conditions on the $\vp$ and $u$-fields. We find that all three eigenvalues $\vp_a$ have to depend only on the radial distance $r$, $\vp_1$ and $\vp_2$ have to be equal, and the complex $u$ field have to correspond to configurations of unity topological charge. Those conditions are given in \rf{varphiansatz} and \rf{q=1ansatz}.  In order to construct the solutions we have to solve therefore just three ordinary differential equations, corresponding to the Euler-Lagrange for the profile function $f\(r\)$ and for the eingenvalues $\vp_1\(r\)$ and $\vp_3\(r\)$, of the matrix $h$. Those equations are solved numerically using the gradient flow method to minimize the static energy of the system.  We perform the simulations for a potential which is quadratic in the $\vp$-fields, i.e., proportional to ${\rm Tr}\,h^2$. Qualitatively the solutions look similar to the self-dual solutions. However, the profile function $f\(r\)$ and the fields $\vp_1\(r\)$ and $\vp_3\(r\)$, decay exponentially faster, at large distances, with the increase of the coupling constants associated to the kinetic and potential terms of the $h$-fields. In addition, $\vp_1\(r\)$ and $\vp_3\(r\)$, grow at the origin with the increase of those same constant constants.

The results we have obtained may shed some light on the structures underlying the self-duality in models of the type \rf{modelintro}. It would be interesting to generalize our results by breaking completely the self-duality, i.e. by not imposing  \rf{halfbpscond}, and construct solutions with topological charges higher than unity, by performing three dimensional numerical simulations to minimize the static energy. That could help to understand better the role of the $h$-fields. In addition, it could help to improve the applications to nuclear matter done in \cite{usfalse}, by performing the breaking of the self-duality with the introduction of kinetic and potential terms for the $h$-fields. \\

\vspace{1cm}

{\bf Acknowledgements:} LAF is supported by Conselho Nacional de Desenvolvimento Cient\'ifico e Tecnol\'ogico - CNPq (contract 308894/2018-9), and Funda\c c\~ao de Amparo \`a Pesquisa do Estado de S\~ao Paulo - FAPESP (contract 2022/00808-7) and LRL is supported by CAPES scholarship. The authors are grateful for the discussions on numerical methods with Wagner Schlindwein. 

\vspace{1cm}


\appendix 

 \section{The solutions of \rf{funnyrel} }
\label{sec:algebraicsol}
\setcounter{equation}{0}

Let us introduce the quantities 
\br
\alpha_1=m_0\,e_0\,\vp_1\,\sqrt{\frac{\omega_1}{\omega_2\,\omega_3}};\quad\;
\alpha_2=m_0\,e_0\,\vp_2\,\sqrt{\frac{\omega_2}{\omega_1\,\omega_3}};\quad\;
\alpha_3=m_0\,e_0\,\vp_3\,\sqrt{\frac{\omega_3}{\omega_1\,\omega_2}};\quad\;
\beta=\frac{\Lambda}{m_0\,e_0}
\nonumber\\
\lab{alphabetadef}
\er
Then the equations \rf{funnyrel} can be written as 
\br
\alpha_1+\frac{1}{\alpha_2}+\frac{1}{\alpha_3}=\beta;\qquad\qquad
\alpha_2+\frac{1}{\alpha_1}+\frac{1}{\alpha_3}=\beta;\qquad\qquad
\alpha_3+\frac{1}{\alpha_1}+\frac{1}{\alpha_2}=\beta
\lab{funnyrel1}
\er
Subtracting the equations \rf{funnyrel1} in pairs we observe that
\be
\alpha_1-\frac{1}{\alpha_1}=\alpha_2-\frac{1}{\alpha_2}=\alpha_3-\frac{1}{\alpha_3}
\lab{funnyrel2}
\ee
In addition, we can write \rf{funnyrel1} as
\be
\beta\,\alpha_1\,\alpha_2-\alpha_1-\alpha_2=\beta\,\alpha_2\,\alpha_3-\alpha_2-\alpha_3=\beta\,\alpha_1\,\alpha_3-\alpha_1-\alpha_3=\alpha_1\,\alpha_2\,\alpha_3
\lab{funnyrel3}
\ee
Again, subtracting the relations \rf{funnyrel3} in pairs we get that
\be
\(\beta\,\alpha_1-1\)\(\alpha_2-\alpha_3\)=0;\qquad
\(\beta\,\alpha_2-1\)\(\alpha_1-\alpha_3\)=0;\qquad
\(\beta\,\alpha_3-1\)\(\alpha_1-\alpha_2\)=0
\lab{funnyrel4}
\ee
Such equations have three types of solutions:
\begin{enumerate}
\item If we take $\alpha_1=\alpha_2=1/\beta$, then all three equations \rf{funnyrel4} are satisfied. Then \rf{funnyrel3} imposes that $\alpha_3=-\beta$. Doing cyclic permutations of the indices we get three solutions
\br
\alpha_1=\alpha_2&=&\frac{1}{\beta}\qquad {\rm and}\qquad \alpha_3=-\beta
\nonumber\\
\alpha_1=\alpha_3&=&\frac{1}{\beta}\qquad {\rm and}\qquad \alpha_2=-\beta
\lab{sol1}\\
\alpha_2=\alpha_3&=&\frac{1}{\beta}\qquad {\rm and}\qquad \alpha_1=-\beta
\nonumber
\er
\item By taking and three $\alpha_a$'s equal to $1/\beta$ we solve  all three equations \rf{funnyrel4}. Then \rf{funnyrel3} imposes that $\beta^2=-1$. So we get the solution
\be
\alpha_1=\alpha_2=\alpha_3=\pm i\qquad\qquad\qquad \beta=\mp i
\lab{sol2}
\ee
which is a particular case of \rf{sol1}.
\item Finally by taking $\alpha_1=\alpha_2=\alpha_3\equiv\alpha$ we solve all three equations \rf{funnyrel4}. Then \rf{funnyrel3} leads to 
\be
\alpha\(\alpha^2-\beta\,\alpha+ 2\)=0
\ee
 The solution $\alpha=0$ should be discarded since from \rf{alphabetadef}, it would imply $\vp_a=0$, and so a vanishing $h$ matrix. Therefore, we get two solutions
\be
\alpha_1=\alpha_2=\alpha_3=\frac{1}{2}\(\beta\pm \sqrt{\beta^2-8}\)=\sqrt{2}\(\gamma\pm\sqrt{\gamma^2-1}\);\qquad\qquad \beta=2\sqrt{2}\,\gamma
\ee
Note that, if we consider $\gamma$ real, we have 
\br
\gamma\geq 1\quad &\rightarrow&\quad\gamma+\sqrt{\gamma^2-1}\geq 1\quad {\rm and} \quad0\leq\gamma-\sqrt{\gamma^2-1}\leq 1
\nonumber\\
\gamma\leq -1\quad &\rightarrow&\quad \gamma-\sqrt{\gamma^2-1}\leq -1\quad {\rm and} \quad
-1\leq\gamma+\sqrt{\gamma^2-1}\leq 0
\nonumber
\er
Therefore, $\alpha_a$ can take any non-negative real  value when $\gamma\geq 1$, and any non-positive real value when $\gamma\leq -1$. For $-1\leq \gamma\leq 1$ we write $\gamma=\cos\theta$, with $0\leq \theta\leq \pi$. Then
\be
\gamma+\sqrt{\gamma^2-1}=e^{i\theta};\qquad {\rm and}\qquad \gamma-\sqrt{\gamma^2-1}=e^{-i\theta}
\ee
Therefore, $\alpha_1=\alpha_2=\alpha_3=\sqrt{2}\, e^{i\theta}$, with $0\leq \theta\leq 2\pi$. If we consider $\gamma$ complex, then the $\alpha_a$'s can in principle be any complex number.  

\end{enumerate}

As argued in \rf{positivetaueigen}, all three eigenvalues $\omega_a$ of the $\tau$-matrix are non negative. Therefore, the solutions \rf{sol1} impose that, if $\beta>0$,  one eigenvalue $\vp_a$ of the $h$-matrix is negative, and,  if $\beta<0$, that two eigenvalues are negative. That implies that the energy $E_1$ is not positive definite. On the hand, the solutions \rf{sol2} imply that the eigenvalues $\vp_a$ of the $h$-matrix are pure imaginary and so $E_1$ is pure imaginary too. 

\section{The rational maps that leads to $q=const.$}
\label{app:proof}
\setcounter{equation}{0}

In this section we will proof that the only rational map \rf{rationalmap} for which the functional $q\(w,\,\bar{w}\)$, defined in \rf{qdef}, is constant corresponds to \rf{q=1ansatz}. Using \rf{rationalmap} in \rf{qdef} we obtain
\be
q\(w,\,\bar{w}\) = A\(w,\,\bar{w}\)^2\,\mid W(w) \mid^2 \,\geq 0,\quad{\rm with}\qquad A\(w,\,\bar{w}\)\equiv \frac{1+\w2}{\mid p_1\mid^2+\mid p_2\mid^2},\lab{lambdan}
\ee
where we introduced the Wronskian $W\equiv p_2\,d_w p_1-p_1\,d_w p_2$. The topological degree $n$ of the $u$-field corresponds exactly with the highest power of $w$ in either of the polynomials $p_1$ and $p_2$, as mentioned in the section \ref{sec:quasisd}, and so $n\geq 1$. Due its definition, if both $p_1$ and $p_2$ are polynomials of degree $n$, then the term of order $n(n-1)$ of the Wronskian will vanishes, and therefore $W$ is a polynomial of degree $n(n-2)$ or less. In additional, the denominator of \rf{lambdan} satisfies $\mid p_1\mid^2+\mid p_2\mid^2 >0$ for all values of $w$ and $\bar{w}$, since  $p_1$ and $p_2$ has no common roots. So, it follows from \rf{lambdan} that $q\(w,\,\bar{w}\)$ vanishes only for the values of $w$ that corresponds with the roots of the Wronskian. Therefore, if $W$ has roots, then $q$ cannot be constant. On the other hand, if the polynomial $W$ has no root, then it must be constant, i.e. the rational map \rf{rationalmap} must satisfy the condition 
\be W(w) = const. \lab{const1}\ee
However, the constant of \rf{const1} cannot be zero, otherwise $q$ will vanishes and  the same goes $n$, which is written in the integral representation in \rf{n}. So it follows from \rf{const1} and \rf{lambdan} that if $q=const.$. Therefore, the function $A$, which is positive and finite due the definition \rf{lambdan}, must also be constant. 

Suppose that $m_1$ and $m_2$ are respectively the degrees of $p_1(w)$ and $p_2(w)$, and let us define $m\equiv {\max }\,\(m_1,\,m_2\)$. Since the algebraic degree of $u$ is equal to $m$, i.e. $m=n$, we get from \rf{lambdan} that 
\be A \sim \mid w \mid^{-2\,\(n-1\)} \qquad \quad {\rm for } \qquad \quad \mid w \mid \gg 1 .\ee
Therefore, the function $A$ can not be constant for every value of $n\geq 2$. Since $q$ is symmetric by exchange $p_1 \leftrightarrow p_2$, the most general rational map $u$ that can be considered with $n=1$ is given by 
\be p_1=\beta\,\(w-a\) \qquad\qquad {\rm and}\qquad\qquad p_2=\(w-b\)^{l},\lab{ansatzpq}\ee
where $l=0,\,1$, the parameters $a,\,b,\,\beta$ are complex numbers and $\beta \neq 0$. In addition, since $p_1$ and $p_2$ do not have common roots, so $a\neq b$ for $l=1$. The quantity $A\(w,\,\bar{w}\)$ of \rf{lambdan} is the ratio between two polynomials in $w$ and $\bar{w}$, with crossed terms. So, since $A$ is constant we can write \rf{lambdan} as a polynomial equation and we must consider two distinct cases: 
\begin{enumerate}

\item \label{item11} The rational map $u(w)=p_1(w)/p_2(w)$ constructed by the relation \rf{ansatzpq} with $l=0$. It then follows from \rf{lambdan} and \rf{ansatzpq} that $A$ will be constant if and only if the following polynomial equation is satisfied 
\be 1 + w\,\bar{w} =A\, \( 1+\mid \beta\mid^2\,\mid a\mid^2+\mid \beta\mid^2 \(w\,\bar{w} -a\,\bar{w}-\bar{a}\,w\) \)\lab{l=0}\ee
Note that the quantities $w$ and $\bar{w}$ are independent. It results from \rf{l=0} that $a=0$, $\mid \beta \mid =1$ and $A=1$, and therefore  \be u=e^{i\,\alpha}\,w,\qquad\qquad \forall \alpha \in [0,\,2\pi),\lab{usolv}\ee 
with the phase $\alpha$ being constant in the physical space. Such a phase was already expected since the function \rf{qdef} has a $U(1)$ global symmetry due its invariance by the transformation $u\rightarrow e^{i\,\alpha}\,u$ and ${\bar u}\rightarrow e^{-i\,\alpha}\,\bar{u}$. Note that for the rational map \rf{usolv} the Wronskian is the phase itself, i.e. $W=e^{i\,\alpha}$, and therefore satisfies the condition \rf{const1}. In addition, due to the definition \rf{lambdan}, the rational map \rf{usolv} implies $q=1$. 
\item \label{item111} The rational map $u(w)=p_1(w)/p_2(w)$ constructed by the relation \rf{ansatzpq} with $l=1$. It then follows from \rf{lambdan} and \rf{ansatzpq} that $A$ will be constant if and only if the following polynomial equation is satisfied  
\br 1 + w\,\bar{w} &=& A\, \left[ \(\mid b \mid^2+\mid \beta\mid^2\,\mid a\mid^2\)+\(1+\mid \beta\mid^2\)\,w\,\bar{w} \right. \nonumber\\ &&\left. -\(b+a\,\mid \beta \mid^2\)\bar{w}-\(\bar{b}+\bar{a}\,\mid\beta\mid^2\)\,w\right],\lab{l=1}\er
where $a\neq b$, which can be written also as
\br 1&=&A\,\(1+\mid \beta \mid^2\) \nonumber\\
    0&=& \mid \beta\mid^2\,a+b \lab{system} \\
    1&=& A\(\mid \beta\mid^2\,\mid a\mid^2+\mid b\mid^2\)
\er
Note that since $a,\,b,\,\beta$ are complex numbers and $A$ is a real number, then the system of algebraic equations \rf{system} have only four real equations for fixing seven real variables. The first and second line of \rf{system} leads to $A=\frac{1}{1+\mid \beta\mid^2}$ and $b=-a\,\mid \beta \mid^2$, respectively. Using such relations, the third line of \rf{system} leads to $\mid a \mid = \mid \beta \mid^{-1}$, or equivalently $\mid b \mid = \mid\beta\mid$. Writting $a$ in the polar form $a=\mid \beta\mid^{-1}\,e^{i\,\alpha}$ we so have that
\be 
p_1 = \beta \,\(w-\mid \beta\mid^{-1}\,e^{i\,\alpha}\);\qquad p_2=w+\mid \beta\mid\,e^{i\,\alpha};\qquad A=\frac{1}{1+\mid\beta\mid^2} \lab{end}
\ee
with the phase $\alpha$ being again a constant in the physical space. Note that due to \rf{end} the polynomials $p_1$ and $p_2$ do not have commum roots and the Wronskian is a complex constant of non-zero modulus, i.e. $W=\frac{\beta}{\mid\beta\mid}\,e^{i\,\alpha}\,\(1+\mid \beta\mid^2\)$. However, due to \rf{lambdan} all the rational maps of the form \rf{end} also lead to $q=1$.
\end{enumerate}

It is concluded that only the rational maps \rf{usolv} and \rf{end} satisfies the condition $q=const.$, and for such maps this constant is determined and we obtain $q=1$. 

\section{The gradient flow method applied to minimize the static energy \rf{modelenergy2}}
\label{app:numerics}
\setcounter{equation}{0}

On this appendix we will discuss the numerical method used in the section \ref{sec:numerics}. We use the gradient flow method with adaptive step size to minimize the static energy \rf{energiessum} and to get the solutions of the equations of motion \rf{deltaef}, \rf{deltaep1} and \rf{deltaep3} with $Q=1$. The range of $r$ considered is $[0,\, r_{\rm max}]$, where the value of $r_{\rm max}$, and so the size of the lattice, can depend of $\sigma_1$ and $\sigma_2$ and is chosen to ensure that $0<\vp_a\(r_{\rm max.}\)<8\times 10^{-7}$, $f\(r_{\rm max.}\)< 4\times 10^{-8}$. The interval between neighboring points is $\Delta r =0.005$ and the grid has $p=r_{{\rm max.}}/\Delta r$ points parametrized by an integer $k$, where we replace $r \rightarrow \Delta r\, k$. 

We use the discrete version of the equations of motion \rf{deltaef}, \rf{deltaep1} and \rf{deltaep3} and the energies \rf{staticenergies}, where the first an second derivatives are given by the central formula of fourth order and the integrals are computed with the trapezoidal rule. In particular, at $k=1$ and $k=p-2$ the derivatives are calculated with the central formula of second order and at $k=p-1$ we use the first order backward difference formula.\footnote{ Because of the Jacobian term $r^2$ only the kinect term of $h$ gives a non-trivial contribution to the energy at $k=0$, given by  $2\frac{r_{\rm max.}}{p}(\vp_{1,0}-\vp_{3,0})^2$, which by \rf{expansion0} can only be non-zero for a field configuration that is not a static solution of the equations \rf{deltaef}-\rf{deltaep3}. Therefore, we do not need computing any derivative at $k=0$.}

The gradient flow method will start with a field configuration with finite energy, called seed configuration, chosen as the discrete version of the self-dual configuration
\be f\(\zeta\)=8\arctan \(e^{-\zeta}\); \qquad\quad \vp_1\(\zeta\)=-f^\prime;\qquad\quad \vp_3\(\zeta\)=\widehat{\vp}_3\(\zeta\)=-\frac{4\sin^2 \(f/2\)}{\zeta^2 f'} \lab{seed}\ee
that satisfies \rf{expansion0}, which will have its fields successively modifies on each discrete point $j=2,...,\,p-1$ and $k=1,...,\,p-1$ of the grid by
\be f_j \rightarrow f_j - \Delta \alpha_f \Delta E_{f,j};\quad\qquad \vp_{b,k}\rightarrow \vp_{b,k}-\Delta \alpha_{\vp_b}\Delta E_{\vp_b , k};\quad b=1,\,3 \lab{change}\ee
where $\Delta \alpha_a$ represent the finite step size. The fields values $f_1$, $\vp_{1,0}$ and $\vp_{3,0}$ can be estimated from the neighbors points, once \rf{change} is done, using 
\be f_1= \frac{f_0+f_2}{2};\qquad\qquad\qquad\vp_{a,0}=\vp_{a,4}-2\(\vp_{a,3}-\vp_{a,1}\) \lab{k=0}\ee
The first equality of \rf{k=0} is obtained using $f''(0)=0$ where the second derivative is given by the second order forward formula. The second equality of \rf{k=0} is obtained taking the equality between the central formula of second and fourth order to the first derivative.\footnote{The equation \rf{deltaef} can be very sensible at $k=1$, where the derivatives are calculated by the second-order central formula, in the sense that small variations in $f$ can lead to large variations of $\Delta E_f$, which in turn may be at odds with its neighboring values in the grid. For example, consider the particular case $\sigma_2=0$ and take a self-dual field configuration that satisfies \rf{deltaef} as a seed configuration, such as \rf{seed}, i.e. for which we have exactly $\Delta E_f=0$. The numerical value of $\Delta E_{f,\,1}$, obtained without replacing \rf{seed} in \rf{deltaef}, can  becomes non-negligible differing significantly from $\Delta E_{f,\,2 }$. So, it may be preferable to use the expression \rf{k=0} to calculate $f_1$ instead the central formula. In addition, since $\Delta E_{f}$ can be very sensitive by small variations of $f$, a very small value of the step size $\Delta \alpha_f$ is used. Finally, the adaptive step size $\Delta \alpha_f$ decreases and the field configuration update is not accepted when the energy \rf{energiessum} grows or $\Delta_f E $ deforms drastically, which can avoid discontinuity problems in the function $\Delta E_f$.} The method  ends when the maximum value of each $|\Delta E_{f} |,\, |\Delta E_{\vp_1}|$ and $|\Delta E_{\vp_3}|$ on the grid, restricted to the points considered in \rf{change}, is smaller than and the Derrick$<4\times 10^{-4}$.

Once we obtain the solutions of \rf{deltaef}-\rf{deltaep3}, where the fields $f(r)$ and $\vp_a(r)$ are monotonic, we can compute the thickness $t_f$, defined as the value of $r$ that satisfies $f(r)=f(0)/2$. First, we get the value $r_p$ which is the numerical value of $r$ on the grid that minimizes the function $|f(0)/2 - f(r)|$, and so by definition $\mid t_f-r_p \mid \leq \Delta r$. So, the Taylor expansion of $f(r)$ at $r_p$ valued for $t=t_f$ becomes $f(t_f)=f(r_p)+f'(r_p)\,\(t_f-r_p\)+\mathcal{O}(\(\Delta r\)^2)$ and then 
\be t_f = r_p+\frac{f(0)/2-f(r_p)}{f'\(r_p\)}+\mathcal{O}\(\(\Delta r\)^2\) \lab{thick}\ee
We use \rf{thick} to compute $t_f$ in first order of $\Delta r$, and the same follows for get the thickness of the $\vp_a$-fields. On the section \ref{app:numerics} we present in the Table \ref{table2} the values of the thickness for each of the fields and for each of the values of $\sigma_1$ and $\sigma_2$ of the Table \ref{table1}.


\begin{thebibliography}{10}

\bibitem{genbps}
C.~Adam, L.~A. Ferreira, E.~da~Hora, A.~Wereszczynski, and W.~J. Zakrzewski.
\newblock {Some aspects of self-duality and generalised BPS theories}.
\newblock {\em JHEP}, 08:062, 2013.

\bibitem{skyrme1}
T.~H.~R. Skyrme.
\newblock {A Nonlinear field theory}.
\newblock {\em Proc. Roy. Soc. Lond. A}, 260:127--138, 1961.

\bibitem{skyrme2}
T.~H.~R. Skyrme.
\newblock {A Unified Field Theory of Mesons and Baryons}.
\newblock {\em Nucl. Phys.}, 31:556--569, 1962.

\bibitem{mantonruback}
N.~S. Manton and P.~J. Ruback.
\newblock {Skyrmions in Flat Space and Curved Space}.
\newblock {\em Phys. Lett. B}, 181:137--140, 1986.

\bibitem{adam1}
C.~Adam, J.~Sanchez-Guillen, and A.~Wereszczynski.
\newblock {A Skyrme-type proposal for baryonic matter}.
\newblock {\em Phys. Lett. B}, 691:105--110, 2010.

\bibitem{adam2}
C.~Adam, J.~Sanchez-Guillen, and A.~Wereszczynski.
\newblock {A BPS Skyrme model and baryons at large $N_c$}.
\newblock {\em Phys. Rev. D}, 82:085015, 2010.

\bibitem{adam_prl}
C.~Adam, C.~Naya, J.~Sanchez-Guillen, and A.~Wereszczynski.
\newblock {Bogomol\textquoteright{}nyi-Prasad-Sommerfield Skyrme Model and
  Nuclear Binding Energies}.
\newblock {\em Phys. Rev. Lett.}, 111(23):232501, 2013.

\bibitem{sut_tower}
Paul Sutcliffe.
\newblock {Skyrmions, instantons and holography}.
\newblock {\em JHEP}, 08:019, 2010.

\bibitem{sut_naya_1}
Carlos Naya and Paul Sutcliffe.
\newblock {Skyrmions in models with pions and rho mesons}.
\newblock {\em JHEP}, 05:174, 2018.

\bibitem{sut_naya_2}
Carlos Naya and Paul Sutcliffe.
\newblock {Skyrmions and clustering in light nuclei}.
\newblock {\em Phys. Rev. Lett.}, 121(23):232002, 2018.

\bibitem{bpswojtek}
L.~A. Ferreira and Wojtek~J. Zakrzewski.
\newblock {A Skyrme-like model with an exact BPS bound}.
\newblock {\em JHEP}, 09:097, 2013.

\bibitem{bpsshnir}
L.~A. Ferreira and Ya. Shnir.
\newblock {Exact Self-Dual Skyrmions}.
\newblock {\em Phys. Lett. B}, 772:621--627, 2017.

\bibitem{laf2017}
L.~A. Ferreira.
\newblock {Exact self-duality in a modified Skyrme model}.
\newblock {\em JHEP}, 07:039, 2017.

\bibitem{us}
L.~A. Ferreira and L.~R. Livramento.
\newblock {Self-Duality in the Context of the Skyrme Model}.
\newblock {\em JHEP}, 09:031, 2020.

\bibitem{usfalse}
L.~A. Ferreira and L.~R. Livramento.
\newblock {A False Vacuum Skyrme Model for Nuclear Matter}.
\newblock {\em arXiv}, 2106.13335 [hep-th], 2021.

\bibitem{mantonbook}
N.~S. Manton and P.~Sutcliffe.
\newblock {\em {Topological solitons}}.
\newblock Cambridge Monographs on Mathematical Physics. Cambridge University
  Press, 2004.

\bibitem{rational1}
Conor~J. Houghton, Nicholas~S. Manton, and Paul~M. Sutcliffe.
\newblock Rational maps, monopoles and skyrmions.
\newblock {\em Nuclear Physics B}, 510(3):507--537, 1998.

\bibitem{rational2}
Richard~A. Battye and Paul~M. Sutcliffe.
\newblock Skyrmions, fullerenes and rational maps.
\newblock {\em Reviews in Mathematical Physics}, 14(01):29--85, 2002.

\bibitem{skyrmejoaq}
L.A. Ferreira and J.Sánchez Guillén.
\newblock Infinite symmetries in the skyrme model.
\newblock {\em Physics Letters B}, 504(1):195--200, 2001.

\bibitem{derrick}
G.~H. Derrick.
\newblock {Comments on nonlinear wave equations as models for elementary
  particles}.
\newblock {\em Journal of Mathematical Physics}, 5(9):1252--1254, 1964.
\newblock DOI: 10.1063/1.1704233.

\bibitem{coleman2}
Sidney~R. Coleman, V.~Glaser, and Andre Martin.
\newblock {Action minima among solutions to a class of euclidean scalar field
  equations}.
\newblock {\em Communications in Mathematical Physics}, 58(2):211--221, 1978.
\newblock DOI: 10.1007/BF01609421.

\end{thebibliography}
\end{document}